\documentclass[a4paper,10pt]{article} 

\onecolumn
\def\sloppy{\tolerance=100000\hfuzz=\maxdimen \vfuzz=\maxdimen}
\usepackage{graphicx}
\usepackage{amssymb}
\usepackage{amsmath}
\usepackage{xfrac}

\newcommand{\rme}{\mathrm{e}}
\newcommand{\rmi}{\mathrm{i}}
\newcommand{\rmd}{\mathrm{d}}

\vspace{-8ex}\date{}

\begin{document}
\sloppy

\title{\Large\bf Structures associated with the Borromean rings\\complement in the Poincar\'e ball} 

\author{Anton A. Nazarenko${}^\dag$ and A.V. Nazarenko${}^\ddag$\\ 
{\small\it ${}^\dag$Faculty of Mechanics and Mathematics, Taras Shevchenko National University of Kyiv,}\\
{\small\it UA--01601 Kyiv, Ukraine}\\
{\small anton.nazarenko02@gmail.com}\\
{\small\it ${}^\ddag$Bogolyubov Institute for Theoretical Physics of NAS of Ukraine,}\\
{\small\it UA--03143 Kyiv, Ukraine}\\
{\small nazarenko@bitp.kyiv.ua}}

\maketitle

\begin{abstract}
Guided by physical needs, we deal with the rotationally isotropic Poincar\'e
ball, when considering the complement of Borromean rings embedded in it.
We consistently describe the geometry of the complement and realize the
fundamental group as isometry subgroup in three dimensions. Applying this
realization, we reveal normal stochastization and multifractal behavior
within the examined model of directed random walks on the rooted Cayley
tree, whose six-branch graphs are associated with dendritic polymers. According to 
Penner, we construct the Teichm\"uller space of the decorated ideal octahedral
surface related to the quotient space of the fundamental group action. Using
the conformality of decoration, we define six moduli and the mapping class
group generated by cyclic permutations of the ideal vertices. Intending
to quantize the geometric area, we state the connection between the induced
geometry and the sine-Gordon model. Due to such a correspondence we obtain
the differential two-form in the cotangent bundle.

{\it Keywords:} {Borromean rings complement; fundamental group; Cayley tree;
random walk; decorated Teichm\"uller space; sine-Gordon equation}
\end{abstract}

\section{Introduction}

Numerous physical problems require their formulation in isotropic three-dimensional
space, an example of which is the Poincar\'e unit ball model with its own group of
isometries, including three-dimensional rotations. The hyperbolicity of the latter
implies the consideration of geometric structures with a negative Euler characteristic,
which determines the number of topological degrees of freedom, also used for physical
modeling. Here we focus on the complement of Borromean rings (BRC), which represent
the simplest Brunnian link~\cite{Th78}, and some of the structures associated with it.
In studying them, we adhere to the chosen space model, although the
often used Klein model is mathematically convenient due to its group of $SL(2,\mathbb{C})$
isometries~\cite{W78,Mat06,Abe13}. 

Dealing with the basic homotopy groups of the link and decorated Teichm\"uller
space~\cite{Penn87}, we goal to provide a framework suitable for further use in
physics\footnote{We are forced to omit here a detailed review of many works
devoted to the study and use of Borromean rings, but note some areas of their
appearance.}. It is due to associating the Borromean rings with quantum
entanglement~\cite{Kauff02,Iq24}, Efimov trimers~\cite{Grimm06,BrH06},
polymers~\cite{Ch04}, as well as for the development of quantum geometry, i.e.
finding the spectra of quantized geometric characteristics~\cite{Rovelli14}
and/or using the quantum groups~\cite{Kas}. The last point, in our opinion, is of
conceptual importance, and therefore the research begun here looks promising.

Having implemented the fundamental group generated by three-dimensional parabolic
generators, the problem of symmetrization of functions with respect to the group
operation naturally arises~\cite{Mat06}. In this regard, we consider the Cayley
tree embedded in the Poincar\'e ball and rooted at the origin to formulate
the partition function of the directed random walk model related to polymer
physics~\cite{dp21}. Defining this function up to $N$th generation, we expect rapid
stochastization of the terms of the Poincar\'e-type series and the revelation of 
multifractality. The sought multifractal exponents are appropriate
for comparing different models and allow interpretation in the spirit of
statistical physics~\cite{Fed}. Besides, resorting to Markov chains to compute
characteristics is similar to using a mean field approximation.

We also focus on the deformations (classes) of conformal structures induced on
the surface of a regular hyperbolic octahedron, the ideal vertices of which are
fixed by the parabolic generators. Using the realized group as a marking, we
involve the decorated Teichm\"uller space and its mapping class group operating
through vertex permutations. Decorating implies the inclusion of horospheres 
centered at the vertices and obtaining curves of their intersection with octahedron
faces, always orthogonal to the corresponding edges of the octahedron~\cite{Penn87}.
Then conformality is ensured by the conservation of right angles regardless of
the size of the horospheres~\cite{BPS15,GL18}, when the horosphere size is changed
by hyperbolic boost~\cite{Penn87}.

In order to determine a differential two-form needed for further geometry quantization,
we search a connection with a suitable dynamical model. It can be realized
by relating the angular size of each intersection curve and the hyperbolic distance
from it to the origin in order to reveal the kink of the sine-Gordon model~\cite{Man04}.
This should allow us to induce the differential two-form within the Hamiltonian formalism,
as well as construct algebra of geometric quantities in the future. We admit the 
applicability of this strategy to developing quantum geometry~\cite{Naz05,Naz13}
and field theory~\cite{Hurt}.

The paper is organized as follows. In Sec.~\ref{Sec2}, we describe the geometry
of the complement of Borromean rings in the Poincar\'e ball and implement the
fundamental group as its isometry subgroup. Multifractal exponents of a
directed random walk model on the Cayley tree are studied in Sec.~\ref{Sec3} using
numerical and approximate methods. In Sec.~\ref{Sec4}, we study the structure of
the decorated Teichm\"uller space and its mapping class group. Introducing
the moduli, we connect the induced geometry and the sine-Gordon model. We finish
our considerations with the Discussion.

\section{\label{Sec2}Geometry and symmetry of the Borromean rings complement}

\subsection{Spaces and their isometries}

Let us define the spaces and their symmetries that we shall use. According to Thurston's
arguments~\cite{Th78}, good knots and links induce a hyperbolic structure in
the three-dimensional space where they naturally exist. Denote by ${\cal H}_3$
the three-dimensional manifold embedded in $\mathbb{R}^3$ and equipped with a hyperbolic
metric of constant negative curvature. It is useful to split ${\cal H}_3$ into
hyperplanes and complexify them:
\begin{equation}
{\cal H}_3\ni x=(x_1,x_2,x_3) \mapsto (x_1+\rmi x_2, x_3)=(z,t)\in \mathbb{C}\times\mathbb{R}.
\end{equation}

Thus, fixing the hyperplane $t=0$, we require that the resulting two-dimensional mani\-fold
${\cal H}_2\ni z\simeq (z,0)$ inherits the hyperbolic structure ($\simeq$ denotes
isomorphic equivalence). This (orthogonal) projection becomes apparent by considering
the two Poincar\'e models: unit ball
 $\mathbb{B}=\{x\in\mathbb{R}^3|\,\|x\|_{\mathbb{B}}^2=x_1^2+x^2_2+x^2_3<1\}\simeq{\cal H}_3$
and unit disc
$\mathbb{D}=\{z\in\mathbb{C}|\,\|x\|_\mathbb{D}^2=|z|^2<1\}\simeq{\cal H}_2$.
Their infinitesimal intervals $\rmd s^2_{\mathbb{B},\mathbb{D}}$ and distances
${\rm dist}_{\mathbb{B},\mathbb{D}}$ are given by the common formulas:
\begin{eqnarray}
&&\rmd s^2_{\mathbb{B},\mathbb{D}}=4\,\frac{\|\rmd x\|_{\mathbb{B},\mathbb{D}}^2}
{(1-\|x\|_{\mathbb{B},\mathbb{D}}^2)^2},\\
&&{\rm dist}_{\mathbb{B},\mathbb{D}}(x,y)=2\,{\rm arcsinh}\,\frac{\|x-y\|_{\mathbb{B},\mathbb{D}}}
{\sqrt{(1-\|x\|_{\mathbb{B},\mathbb{D}}^2) (1-\|y\|_{\mathbb{B},\mathbb{D}}^2)}}.
\end{eqnarray}
These are also generalized for $n$ dimensions and lead to the Gaussian curvature $K=-1$.

Geodesics in $\mathbb{B}$ and $\mathbb{D}$ are either diameters passing through
the origin or arcs ortho\-gonally intersecting the boundaries $\partial\mathbb{B}$
and $\partial\mathbb{D}$. To give the geodesic ${\bf r}(\theta)$ connecting two
points with radius-vectors ${\bf r}_1,\,{\bf r}_2\in\mathbb{B}$, we suggest
the parametrization:
\begin{eqnarray}
&&{\bf r}(\theta)={\bf r}_0+({\bf r}_1-{\bf r}_0) \cos{\theta}+
[{\bf n}\times({\bf r}_1-{\bf r}_0)] \sin{\theta},\quad {\bf r}(0)={\bf r}_1;\label{geo1}\\
&&{\bf r}_0=\frac{{\bf n}\times{\bf m}}{|{\bf r}_1\times{\bf r}_2|},\quad
{\bf n}=\frac{{\bf r}_1\times{\bf r}_2}{|{\bf r}_1\times{\bf r}_2|},\quad
{\bf m}=c_2 {\bf r}_1-c_1 {\bf r}_2,\quad
c_i=\frac{{\bf r}_i^2+1}{2}.
\nonumber
\end{eqnarray}

In fact, (\ref{geo1}) describes a circular arc with center at ${\bf r}_0$ and
radius $R=|{\bf r}_1-{\bf r}_0|=|{\bf r}_2-{\bf r}_0|=\sqrt{{\bf r}_0^2-1}$;
${\bf n}$ is the normal to the circle plane covering three points ${\bf r}_1$,
${\bf r}_2$, and the origin ${\bf r}=0$ that do not lie on the same straight line.
The equation $|{\bf r}(\theta_2)-{\bf r}_2|=0$ defines the parameter $\theta_2$,
and one has (numerically) that
\begin{equation}
{\rm dist}_{\mathbb{B}}({\bf r}_1,{\bf r}_2)=2R\int_0^{\theta_2}\frac{\rmd\theta}{1-{\bf r}^2(\theta)}.
\end{equation}

Let us first turn to ${\cal H}_2\simeq(\mathbb{D},\rmd s^2_\mathbb{D})$ for our geometry
and group constructions. It implies the use of an orientation-preserving subgroup of
isometries ${\rm Isom}(\mathbb{D})$ that preserve $\rmd s^2_{\mathbb{D}}$, that is,
$PSU(1,1)\simeq SU(1,1)/\{\pm1\}$, whose element $g$ acts freely on $z\in\mathbb{D}$
via a linear-fractional transformation:
\begin{equation}
g[z]=\frac{uz+v}{{\bar v}z+{\bar u}},\quad
g=\left(
\begin{array}{cc}
u & v\\
{\bar v} & {\bar u}
\end{array}
\right),\quad
\det{g}=|u|^2-|v|^2=1.
\end{equation}

Then, having obtained a matrix representation of some group $\Gamma\subset PSU(1,1)$,
we need to extend the action of its generators $g=(g_{i,j})$ up to the ball
$\mathbb{B}$:
\begin{equation}
g[(z,t)]=(z_g(z,t),t_g(z,t)),\qquad
g[(z,0)]=(g[z],0),
\end{equation}
such that ${\rm dist}_{\mathbb{B}}(x,y)={\rm dist}_{\mathbb{B}}(x_g,y_g)$
for $x=(z_1,t_1)$, $y=(z_2,t_2)$, and their images $x_g=(z_g(z_1,t_1),t_g(z_1,t_1))$,
$y_g=(z_g(z_2,t_2),t_g(z_2,t_2))$. In general, this can be done using explicit
formulas, for example, from~\cite{MNY05}.

For our purposes, starting with parabolic generators of $PSU(1,1)$:
\begin{equation}
g=\left(
\begin{array}{cc}
1+\rmi\,a & -\rmi\,a\,\rme^{\rmi\varphi}\\
\rmi\,a\,\rme^{-\rmi\varphi} & 1-\rmi\,a
\end{array}
\right),\quad
g[\rme^{\rmi\varphi}]=\rme^{\rmi\varphi},\quad
a,\,\varphi\in\mathbb{R},
\end{equation}
when $g^n$ is simply obtained by means of replacement $a\mapsto na$,
we write down the extended action of $g[{\bf r}]$ for
${\bf r}\equiv(x\ y\ t)^\top\in\mathbb{B}$ in terms of the
linear combinations:
\begin{equation}
\xi_1=x \sin{\varphi}-y \cos{\varphi},\quad
\xi_2=y \sin{\varphi}+x \cos{\varphi};\quad
x^2+y^2=\xi_1^2+\xi_2^2.
\end{equation}

Thus, we derive that
\begin{eqnarray}
&&
{\footnotesize
g\left[\left(\begin{array}{c} x\\ y\\ t \end{array}\right)\right]=
\begin{array}{c} 1 \\ \overline{a^2[t^2+(\xi_2-1)^2]+(a\xi_1+1)^2} \end{array}
\left(\begin{array}{c}
 {\tilde x}\\
 {\tilde y}\\
 t
\end{array}\right)}
,\label{gext}\\
&&{\tilde x}=x+(a^2\cos{\varphi}+a\sin{\varphi})[\xi_1^2+(\xi_2-1)^2+t^2]+2a\xi_1\cos{\varphi},
\nonumber\\
&&{\tilde y}=y+(a^2\sin{\varphi}-a\cos{\varphi})[\xi_1^2+(\xi_2-1)^2+t^2]+2a\xi_1\sin{\varphi}.
\nonumber
\end{eqnarray}

It serves to get the group $\Gamma_*\subset{\rm Isom}(\mathbb{B})$ from $\Gamma$. 
Obviously, $\Gamma_*$ based on $\Gamma$ cannot take into account all symmetries of
three-dimensional objects. Therefore, additional generators of ${\rm Isom}(\mathbb{B})$
should be introduced, by operating, for example, with $SO(3)$ rotations due to 
the spherical symmetry supposed.

Indeed, the Rodrigues' formula~\cite{Rod} allows us to present the rotation of ${\bf r}$
by an angle $\varphi$ around the direction along the unit vector ${\bf n}$ as
\begin{equation}\label{RRot}
R_{{\bf n},\varphi}[{\bf r}]={\bf r}\cos{\varphi}+{\bf n} ({\bf n}\cdot{\bf r}) (1-\cos{\varphi})
+[{\bf n}\times{\bf r}] \sin{\varphi}.
\end{equation}
It is such that $R^{-1}_{{\bf n},\varphi}=R_{{\bf n},-\varphi}$;
$R_{{\bf m},\varphi}=R\, R_{{\bf n},\varphi}\, R^{-1}$ for ${\bf m}=R[{\bf n}]$
and rotation $R$.

At the end, we also note the HNN group extension~\cite{HNN} and 
the M\"obius transformations in terms of quaternions~\cite{Wil93} (see also
references therein).

\subsection{The Borromean rings complement}

\begin{figure}[htbp]
\begin{center}
\includegraphics[width=5.5cm,angle=0]{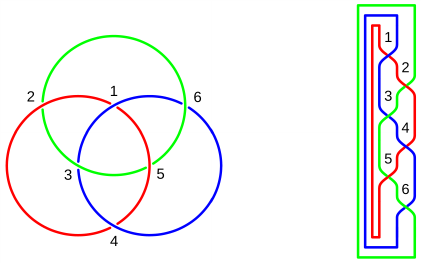}
\end{center}
\caption{\label{fig1}\small Left panel: Sketch of Borromean rings. There are
{\it seven} intrinsic triangles formed by intersections, namely (123), (135),
(156), (126), (234), (345), (456), and {\it one} extrinsic triangle (246).
Right panel: The closed braid with a top-to-bottom direction. There is a
group relation $(\sigma_2\sigma_1^{-1})^3=1$ in terms of crossings $\sigma_1$
(between red and blue strands) and $\sigma_2$ (between blue and green ones).}
\end{figure}

Let us consider the link which is formed by Borromean rings and sketched
in Fig.~\ref{fig1}(left), and define the symmetry of its complement. This link is
isotopically equivalent to the braid in Fig.~\ref{fig1}(right), where the ends of
each strand are closed.

As was argued by Thurston~\cite{Th78}, the Borromean rings complement (BRC)
is a hyperbolic three-manifold $M$ that has a tessellation consisting of two
ideal regular octahedra. The group $G$ of isometries of $M$ acts freely and
transitively on the set of flags of this tessellation, and it has to be of
the order 48.

\begin{figure}[htbp]
\begin{center}
\includegraphics[height=4.3cm,angle=0]{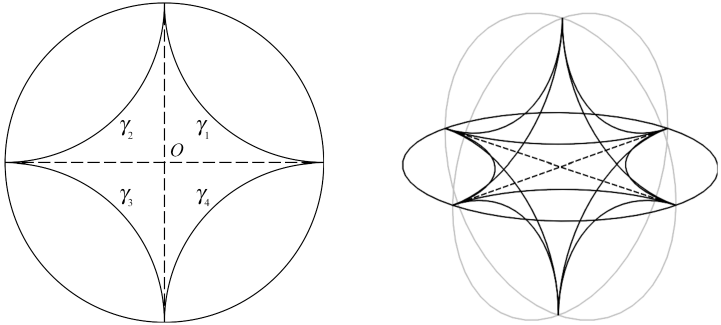}
\end{center}
\caption{\label{fig2}\small Left panel: Basis of a hyperbolic octahedron
with ideal vertices in $\mathbb{D}$ plane. Right panel: Centered hyperbolic
octahedron in $\mathbb{B}$.}
\end{figure}

We start from considering the ideal regular octahedron in Fig.~\ref{fig2}
with ideal vertices at $\pm{\bf i}$, $\pm{\bf j}$, $\pm{\bf k}$, where
$\{{\bf i},{\bf j},{\bf k}\}$ is the standard basis of $\mathbb{R}^3$. Its
edges are the twelve geodesic arcs within the ball, see (\ref{geo1}).

The octahedron basis in $\mathbb{D}$ is stabilized by the parabolic
generators of the group
$\Gamma=\langle h_1, h_2\,|\,(h_1 h_2)^2=-I\rangle$, where
\begin{equation}
h_1=\left(
\begin{array}{cc}
1-\rmi & -1 \\
-1 & 1+\rmi
\end{array}
\right),\qquad
h_2=\left(
\begin{array}{cc}
1-\rmi & -\rmi \\
\rmi & 1+\rmi
\end{array}
\right).
\end{equation}

Defining $h_3=h_2 h_1 h_2^{-1}$ and $h_4=h_1^{-1} h_2 h_1$ so that
$(h_2 h_3)^2=(h_3 h_4)^2=(h_1 h_4)^2=h_4 h_3 h_2 h_1=-I$, every generator
$h_k$ ($k=\overline{1,4}$) fixes the point 
$z_k=\exp{(\rmi\pi k/2)}\in\partial\mathbb{D}$ and determines the mapping
$h_k:\,\gamma_k\to\gamma_{k+1}$ ($\gamma_5=\gamma_1$) in Fig.~\ref{fig2}(left).
As we shall see below, the extensions of $h_k$ connect also the vertical edges
of the octahedra whose bases lie in $\mathbb{D}$.
However, in order to fill the entire ball $\mathbb{B}$ with octahedra,
additional tools are needed. In fact, the group $\Gamma$ should be extended up to
$\Gamma_*$ by including generators (\ref{A1})-(\ref{A8}).

\begin{figure}[htbp]
\begin{center}
\includegraphics[height=3.5cm,angle=0]{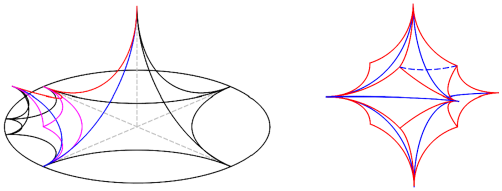} 
\end{center}
\caption{\label{fig3}\small Left panel: Upper parts of two adjacent octahedra.
The colored edges of daughter octahedron bound the tetrahedron, which
should be glued to the nearest face of the central octahedron. The red
and blue edges of tetrahedron in the third quadrant of $\mathbb{D}$ are
identified with the colored edges of the central octahedron by
using $h_2^{-1}$ and $h_3$, respectively. Right panel:
The red-edge dodecahedron covers the central octahedron with blue edges.
This is obtained by taking eight tetrahedra along the equatorial perimeter
of the central octahedron and then gluing them to its eight faces.
Dashed geodesic smoothly connects the nearby finite
vertices of glued tetrahedra.}
\end{figure}

As it was previously argued~\cite{Ad20}, gluing two octahedra together to obtain BRC,
the maximally symmetrical body that can be produced is the rhombic {\it dodecahedron},
which inherits the octahedral symmetry of the order 48. To achieve it, we may cut one 
octahedron into eight tetrahedra with further gluing of each to the eight faces of
another octahedron. We are implementing this by using $\Gamma_*$.

First, acting by the four generators $h_k$ on the (edges of) octahedron in
Fig.~\ref{fig2}(right), four daughter octahedra are obtained with bases in
four quadrants of~$\mathbb{D}$. The new octahedra are geodesic solids in
$(\mathbb{B},\rmd s^2_{\mathbb{B}})$, whose vertical edges (and faces)
with $t\not=0$ are resulted due to (\ref{gext}).
By construction, daughter octahedra have only one edge adjacent to the parent
octahedron, see Fig.~\ref{fig3}(left).

Further, we extract tetrahedra from a daughter octahedron by cutting it with
three hyperplanes of octahedral symmetry. Such an operation is shown in
Fig.~\ref{fig3}(left). Each tetrahedron has right dihedral angles
at the vertex in the center of the octahedron.

In our approach, we glue one face of the adjacent tetrahedron to the face
of the central octahedron, as shown in Fig.~\ref{fig3}(left). This is allowed
due to the geodesic nature of faces and edges of equal areas and lengths.
Seemingly, the procedure of cutting four daughter octahedra with taking two tetrahedra
(in the upper and lower hemisphere) seems easier than decomposing one octahedron
(with basis in $\mathbb{D}$) into eight tetrahedra with subsequent manipulations.
Anyway, this is also admissible using $h_k$. As a result, all procedures lead to
the rhombic dodecahedron (with eight finite vertices at points
$(\pm\sfrac{1}{3}\ \pm\sfrac{1}{3}\ \pm\sfrac{1}{3})^\top$),
which specifies the complement and is shown in Fig.~\ref{fig3}(right).
Thereby, we complete the BRC geometrical description using $\Gamma$ and $\Gamma_*$. 

\subsection{BRC group realization}

Let us now realize the BRC fundamental group $\pi_1({\rm BRC})$ that is
the semi-direct product $\mathbb{Z}_2^3\rtimes {\cal C}_3$, where ${\cal C}_3$
is the third-order cyclic group \cite{RT21,Hoff22}. This is subgroup of the
group $G\simeq\mathbb{Z}_2^3\rtimes S_3$ of three-manifold $M$
with symmetric group $S_3$ acting transitively on the standard basis
of the vector space $\mathbb{Z}_2^3$.

Having specified the parent octahedron Oct as shown in Fig.~\ref{fig2},
the action of $\mathbb{Z}_2^3$ on either octahedra of the tessellation of $M$
corresponds to the reflections in the coordinate hyperplanes of $\mathbb{B}$.
Besides, the elementary group $\mathbb{Z}_2^3$ acts trivially on the set $Y$
of cusps of $M$, while the quotient group $G^{Y}=S_3$ acts transitively on $Y$.

We adopt the Wirtinger representation of 
$\Gamma_{\rm BRC}=\langle g_1, g_2, g_3\,|\,R_1, R_2, R_3\rangle$:
\begin{eqnarray}
&R_1:& (g_2^{-1} g_3 g_2 g_3^{-1}) g_1=g_1 (g_2^{-1} g_3 g_2 g_3^{-1}),
\nonumber\\
&R_2:& (g_3^{-1} g_1 g_3 g_1^{-1}) g_2=g_2 (g_3^{-1} g_1 g_3 g_1^{-1}),
\label{gRel}\\
&R_3:& (g_1^{-1} g_2 g_1 g_2^{-1}) g_3=g_3 (g_1^{-1} g_2 g_1 g_2^{-1}).
\nonumber
\end{eqnarray}
Note that the known $SL(2,\mathbb{C})$ matrix realizations 
are in \cite{W78,Mat06,Abe13}.

We have already stabilized the Oct in $(\mathbb{B},\rmd s^2_{\mathbb{B}})$ by
a subgroup of ${\rm Isom}(\mathbb{B})$ with the {\it functions composition}
as group operation. Now, we are interested in realization
$\varphi:\,\pi_1({\rm BRC})\mapsto {\rm Isom}(\mathbb{B})$  
which implies the mapping $g_i\mapsto\varphi(g_i)$.

We start with the generators of three Abelian subgroups
in terms of ``meridians'' and ``longitudes'' for three fixed points:
\begin{equation}\label{subG}
{\tilde h}^m_1 h_1^n=h_1^n {\tilde h}^m_1,\ \ \
{\tilde h}^m_2 h_2^n=h_2^n {\tilde h}^m_2,\ \ \
{\tilde h}^m_+ h_+^n=h_+^n {\tilde h}^m_+;\ \ \
n,m\in\mathbb{Z}.
\end{equation}
They transform ${\bf r}=(x\ y\ t)^\top\in\mathbb{B}$ so that their
composition reads $h_ih_j[{\bf r}]=h_i[h_j[{\bf r}]]$.

Indeed, every pair $(h_k,{\tilde h}_k)$ is Abelian if the parabolic $h_k$ and
${\tilde h}_k$ have the same fixed point in $\partial\mathbb{B}$.
The needed generators can easily be obtained by using $SO(3)$ rotation (\ref{RRot}).
We have collected the auxiliary generators in Appendix~\ref{App1}.

In principle, (\ref{subG}) means that we have six generators, and we would like
to reduce their number to three by imposing extra relations (\ref{gRel}).

Selecting generators from Appendix~\ref{App1}, we obtain the realization $\Gamma_{\rm BRC}$
operating in $\mathbb{B}$:
\begin{eqnarray}
&&\hspace{-9mm}
{\footnotesize
g_1^n\left[\left(\begin{array}{c} x\\ y\\ t \end{array}\right)\right]=
\begin{array}{c} 1 \\ \overline{n^2[x^2+(y-1)^2]+(nt-1)^2} \end{array}
\left(
\begin{array}{c}
x \\
y-1+n^2[x^2+(y-1)^2]+(nt-1)^2\\
t-n[x^2+(y-1)^2+t^2]
\end{array}
\right),}
\nonumber\\
&&\hspace{-9mm}
{\footnotesize
g_2^n\left[\left(\begin{array}{c} x\\ y\\ t \end{array}\right)\right]=
\begin{array}{c} 1 \\ \overline{n^2[t^2+(x+1)^2]+(ny-1)^2} \end{array}
\left(\begin{array}{c}
 x+1-n^2[t^2+(x+1)^2]-(ny-1)^2\\
 y-n[(x+1)^2+y^2+t^2]\\
 t
\end{array}\right),}
\label{gBRC}\\
&&\hspace{-9mm}
{\footnotesize
g_3^n\left[\left(\begin{array}{c} x\\ y\\ t \end{array}\right)\right]=
\begin{array}{c} 1 \\ \overline{n^2[y^2+(t-1)^2]+(nx+1)^2} \end{array}
\left(\begin{array}{c}
x+n[x^2+y^2+(t-1)^2]\\
y\\
t-1+n^2[y^2+(t-1)^2]+(nx+1)^2
\end{array}
\right),}
\nonumber
\end{eqnarray}
where $n$ is an exponent. 

It is seen that $g_1$, $g_2$, and $g_3$
fix the points $(0\ 1\ 0)^\top$, $(-1\ 0\ 0)^\top$, 
$(0\ 0\ 1)^\top$, respectively. Due to (\ref{RRot}) we have that 
$g_1=R_{{\bf j},\pi/2}\,h_1^{-1}\,R_{{\bf j},-\pi/2}$,
$g_2=h_2$, and $g_3=R_{{\bf i},\pi/2}\,h_1^{-1}\,R_{{\bf i},-\pi/2}$,
while $g_2^{-1}\,g_3\,g_2\,g_3^{-1}=h_1^2$,
$g_3^{-1}\,g_1\,g_3\,g_1^{-1}=
R_{{\bf i},\pi/2}\,h_2^{-2}\,R_{{\bf i},-\pi/2}$, and
$g_1^{-1}\,g_2\,g_1\,g_2^{-1}=R_{{\bf j},\pi/2}\,
h_2^{-2}\,R_{{\bf j},-\pi/2}$.

The easiest way to get other realizations of the group $\Gamma_{\rm BRC}$
is to rotate these generators altogether around the main coordinate axes
as ${\tilde g}_k=R_{{\bf n},\pi/2}\,g_k\,R_{{\bf n},-\pi/2}$, where vector
${\bf n}$ is one of $\{{\bf i},{\bf j},{\bf k}\}$.

\section{\label{Sec3}Cayley tree and multifractality}

One of the direct applications of the group $\Gamma_{\rm BRC}$ is a random walk
model on the generated Cayley tree. Such a statistical model can be used in
dendritic polymer physics~\cite{dp21} for example. It represents an alternative
to two-dimensional models using hyperbolic generators. As an advantage, the
multifractal indices calculated here are easily compared with the others.

\subsection{Construction of the Cayley tree and spectrum analysis}

Using the realization (\ref{gBRC}), let us enumerate the generating set of
$\Gamma_{\rm BRC}$ as
\begin{equation}\label{gami}
\{\gamma_i\,|\,i=\overline{1,6}\}=\{ g_1, g_2, g_3, g_1^{-1}, g_2^{-1}, g_3^{-1}\}.
\end{equation}

By associating the points ${\mathbb B}$ with their radius-vectors and considering
the origin~${\bf 0}$ as the root point, the six-branch Cayley tree of the $N$th
generation is defined as embedded in the ball~${\mathbb B}$ and formed by the set
of $6\times5^{N-1}$ admissible graphs that sequentially connect the vertices:
\begin{equation}
\gamma_{i_1}[{\bf 0}],\ \gamma_{i_2}\gamma_{i_1}[{\bf 0}],\ldots,\
\gamma_{i_N}\gamma_{i_{N-1}}...\gamma_{i_1}[{\bf 0}];\quad
i_t\in\{1,2,3,4,5,6\}.
\end{equation}
An admissible graph with the word $\{i_N,i_{N-1},\ldots,i_1\}$ is allowed by
the conditional probability $p_N(i_N,i_{N-1},\ldots,i_1)=1$ which we define as
\begin{equation}
p_N(i_N,i_{N-1},\ldots,i_1)=\prod\limits_{t=2}^N p_2(i_t,i_{t-1}),\quad
p_2(i_t,i_{t-1})=\left\{
\begin{array}{cc} 0,& |i_t-i_{t-1}|=3\\ 1,& \text{otherwise} \end{array}
\right..
\end{equation}
It is easy to see that $p_N$ prohibits backward steps and makes the graph directed.
Besides, there is the relation:
\begin{equation}
\sum\limits_{i_1=1}^6\ldots\sum\limits_{i_N=1}^6 p_N(i_N,i_{N-1},\ldots,i_1)=v(N),\quad
v(N)=6\times5^{N-1},
\end{equation}
where $v(N)$ is the number of vertices of $N$th generation.
This means that the Cayley tree of the $N$th generation is a rooted pencil of $v(N)$
branches (with common parts).

Strictly speaking, the compositions of the original generators (\ref{gami}) give us
the resulting transformations of all types: parabolic, elliptic, hyperbolic.
It is naively expected that only combinations of the form $\gamma_i^n$ ($n\in{\mathbb Z}$)
and those represented in the brackets of relations (\ref{gRel}) remain parabolic.
These features add interest to the study of the (finite) Cayley tree. In this regard,
we turn to the multifractal analysis.

Our study is based on the $N$th generation partition function of moment order~$q$:
\begin{eqnarray}
&&{\cal Z}_N(q)=\sum\limits_{i_1=1}^6\ldots\sum\limits_{i_N=1}^6 p_N(i_N,i_{N-1},\ldots,i_1)\,
\rme^{q {\cal L}(i_N,i_{N-1},\ldots,i_1)},\label{Z1}\\
&&{\cal L}(i_N,i_{N-1},\ldots,i_1)=\sum\limits_{t=1}^N
{\rm dist}_{\mathbb B}({\bf 0},\gamma_{i_t}\gamma_{i_{t-1}}\ldots\gamma_{i_1}[{\bf 0}]).\label{L1}
\end{eqnarray}

Mathematically, the function ${\cal Z}_N(q)$ is a kind of Poincar\'e series over
a group. Physically, (\ref{Z1}) can be interpreted as the Feynman integral of the
Boltzmann weight over discrete paths, associating the parameter~$q$ with the
inverse temperature $1/T$, which takes positive and negative values. At the same
time, the form of functional ${\cal L}$ should actually provide a second-order phase
transition, the geometric analogue of which is multifractal behavior. Here,
(\ref{L1}) may be considered as the perimetric characteristic of a surface
consisting of adjacent triangles connecting the root point and two points of
different generations. Also note that similar functions were previously used
in \cite{Naz} for a Fuchsian group operating in $({\mathbb D},\rmd s^2_{\mathbb D})$.

From forthcoming analysis it is easily seen that the functional
\begin{equation}
{\cal L}^0(i_N,i_{N-1},\ldots,i_1)=\sum\limits_{t=1}^N
{\rm dist}_{\mathbb B}(\gamma_{i_{t-1}}\ldots\gamma_{i_1}[{\bf 0}],
\gamma_{i_t}\gamma_{i_{t-1}}\ldots\gamma_{i_1}[{\bf 0}])
\end{equation}
does not lead to multifractal behavior because of the vanishing variance of
the ${\cal L}^0$-spectrum. However, such a term may also be involved elsewhere.

Since there are $v(N)$ admissible values of (\ref{L1}), they can be enumerated
with a single index $\zeta$: 
$\{{\cal L}_{\zeta}\,|\,\zeta=\overline{1,v(N)}\}=\{{\cal L}(\{i_t\})\,|\,p_N(\{i_t\})=1\}$,
to reduce ${\cal Z}_N(q)$ to
\begin{equation}\label{Z2}
{\cal Z}_N(q)=\sum\limits_{\zeta=1}^{v(N)} \rme^{q {\cal L}_{\zeta}}; \qquad
{\cal Z}_N(0)=v(N).
\end{equation}

\begin{figure}[htbp]
\begin{center}
\includegraphics[width=12cm,angle=0]{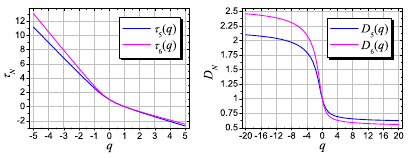}
\end{center}
\vspace*{-4mm}
\caption{\label{fig4}\small Mass spectrum $\tau_N$ (left panel) and
spectrum of fractal dimensions $D_N$ (right panel) as functions of moment order $q$
for different generations $N$.}
\end{figure}

According to the recipe in \cite{Fed}, we define the scaling exponent $\tau_N(q)$,
sometimes called the mass spectrum, and the spectrum of fractal dimensions $D_N(q)$,
associated with the Hausdorff dimensions, up to multiplication by the carrier dimension:
\begin{equation}\label{tD1}
\tau_N(q)=\frac{1}{\ln{v(N)}}\,\left[\ln{{\cal Z}_N(q)}-q\ln{{\cal Z}_N(1)}\right],
\qquad
D_N(q)=\frac{\tau_N(q)}{1-q}.
\end{equation}
A comparison of these functions for different $N$ can be done in Fig.~\ref{fig4}.
As is seen, the slope of the functions to the left of $q=0$ grows as $N$
increases.

Another important characteristic is the spectrum of singularities $f_N(\alpha_N)$,
determined (in parametric form) by using the Legendre transform:
\begin{equation}\label{fa}
f_N(q)=\tau_N(q)-q\,\frac{\rmd\tau_N(q)}{\rmd q},\qquad
\alpha_N(q)=-\frac{\rmd\tau_N(q)}{\rmd q}.
\end{equation}

\begin{figure}[htbp]
\begin{center}
\includegraphics[width=12cm,angle=0]{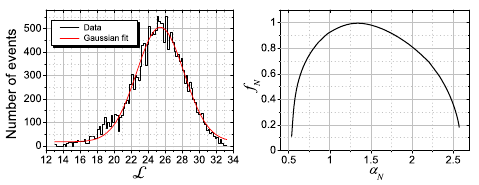}
\end{center}
\vspace*{-4mm}
\caption{\label{fig5}\small Characteristics of the model for $N=6$. Left panel:
The number of ${\cal L}_{\zeta}$ within the intervals $\Delta{\cal L}=0.2$
from ${\cal L}_{\rm min}\simeq13.04$ to ${\cal L}_{\rm max}\simeq33.1$.
Right panel: The spectrum of singularities $f_N(\alpha_N)$; $\alpha_{N,{\rm min}}\simeq0.5353$
and $\alpha_{N,{\rm max}}\simeq2.574$ for $N=6$.}
\end{figure}

By definition, one gets that
\begin{equation}
\alpha_{N,{\rm min}}=\lim\limits_{q\to+\infty} D_N(q),\qquad
\alpha_{N,{\rm max}}=\lim\limits_{q\to-\infty} D_N(q).
\end{equation}

The calculated characteristics indicate normal stochastization with increasing
$N$ in Fig.~\ref{fig5}(left), which also leads to a fairly wide
range of the Lipschitz–H\"older exponent $\alpha_N$ in Fig.~\ref{fig5}(right)
for $N=6$. Firstly, this means that for large $N$ the central limit theorem (CLT)
is applicable~\cite{Knill}. And secondly, a wide range of $\alpha$ shows the absence
of a dominant subset of graphs (fractals) and the need to take into account all graphs,
albeit approximately, when analytically calculating the function~${\cal Z}_N(q)$.
This motivates us to use a Markov chain (random walk) in our studies.

\subsection{Markov chain approximation}

Analyzing (\ref{Z2}), we first assume that the spectrum 
$\{{\cal L}_{\zeta}\,|\,\zeta=\overline{1,v(N)}\}$ is degenerate. In other words, there is
the set $\{{\cal L}_s\,|\,s=\overline{1,S(N)}\}$ with $S(N)\leq v(N)$ of unequal quantities.
Introducing the degeneracy coefficient $w_s$ for each ${\cal L}_s$, we arrive at
\begin{equation}\label{Z3}
{\cal Z}_N(q)=\sum\limits_{s=1}^{S(N)} w_s\,\rme^{q {\cal L}_s}.
\end{equation}

Determining ${\cal L}_{\min}(N)$ and ${\cal L}_{\max}(N)$ such that
${\cal L}_{\min}(N)\leq{\cal L}_s\leq{\cal L}_{\max}(N)$, the next step is
the transition to a continuous limit:
\begin{equation}\label{Z4}
{\cal Z}_N(q)=v(N) \int_{{\cal L}_{\min}(N)}^{{\cal L}_{\max}(N)}
W_N({\cal L})\,\rme^{q {\cal L}}\,\rmd{\cal L}.
\end{equation}

Taking into account observations from Fig.~\ref{fig5}(left), we conclude that
the distribution $W_N({\cal L})$ for large $N$ is approximated by a Gaussian:
\begin{equation}
W_N({\cal L})=A_N\,\exp{\left[-\frac{({\cal L}-\overline{\cal L}_N)^2}{2\sigma_N^2}\right]},
\qquad
\int_{{\cal L}_{\min}(N)}^{{\cal L}_{\max}(N)} W_N({\cal L})\,\rmd{\cal L}=1.
\end{equation}

Simple calculations result in the approximate (Gaussian) expression:
\begin{eqnarray}
&&{\cal Z}^*_N(q)=v(N)\,\frac{c_N(q)}{c_N(0)}\,
\exp{\left(\frac{1}{2}\,q^2\sigma_N^2+q\overline{\cal L}_N\right)},
\label{Zapp}\\
&&c_N(q)={\rm erf}\left(\frac{{\cal L}_{\max}(N)-\overline{\cal L}_N-q\sigma_N^2}{\sqrt{2}\,\sigma_N}\right)
+{\rm erf}\left(\frac{\overline{\cal L}_N+q\sigma_N^2-{\cal L}_{\min}(N)}{\sqrt{2}\,\sigma_N}\right),
\nonumber
\end{eqnarray}
where ${\rm erf}(x)$ is the error function~\cite{AbSt}.

Therefore, the use of CLT allows us to reduce the problem to finding four parameters
${\cal L}_{\min}(N)$, ${\cal L}_{\max}(N)$, $\overline{\cal L}_N$, and $\sigma_N$
that characterize the spectrum $\{{\cal L}_{\zeta}\,|\,\zeta=\overline{1,v(N)}\}$.
In fact, this approximation permits to reproduce the exponent $\tau_N(q)$ for large $N$
and relatively small $|q|$.
Then we are left with the task of obtaining analytical estimates ${\cal L}^*_{\min}(N)$,
${\cal L}^*_{\max}(N)$, $\overline{\cal L}^{\,*}_N$, and $\sigma^*_N$, bypassing
the calculation of the spectrum $\{{\cal L}_{\zeta}\,|\,\zeta=\overline{1,v(N)}\}$.
To implement our program, we appeal to a number of general properties.

Obviously, the spectrum $\{{\cal L}(i_1\ldots i_{N-1},i_N)\,|\,p_N(i_1\ldots i_{N-1},i_N)=1\}$
coincides with $\{{\cal L}(i_N,i_{N-1}\ldots i_1)\,|\,p_N(i_N,i_{N-1}\ldots i_1)=1\}$
for $i_t=\overline{1,6}$ and $t=\overline{1,N}$. Besides, let us recall that
${\rm dist}_{\mathbb B}({\bf 0},{\bf r})={\rm dist}_{\mathbb B}
(\gamma[{\bf 0}],\gamma[{\bf r}])$ and ${\rm dist}_{\mathbb B}({\bf 0},\gamma[{\bf r}])
={\rm dist}_{\mathbb B}(\gamma^{-1}[{\bf 0}],{\bf r})$ for $\gamma\in{\rm Isom}({\mathbb B})$.

There is also the triangle rule for the point set $\{{\bf 0},{\bf r}_{t-1},{\bf r}_t\}$:
\begin{eqnarray}
\cosh{{\rm dist}_{\mathbb B}({\bf 0},{\bf r}_t)}&=&\cosh{{\rm dist}_{\mathbb B}({\bf 0},{\bf r}_{t-1})}\,
\cosh{d_{t,t-1}}\nonumber\\
&&+\sinh{{\rm dist}_{\mathbb B}({\bf 0},{\bf r}_{t-1})}\,\sinh{d_{t,t-1}}\,
\cos{\psi_{t,t-1}},\label{trr}
\end{eqnarray}
where $d_{t,t-1}={\rm dist}_{\mathbb B}({\bf r}_t,{\bf r}_{t-1})$;
$\psi_{t,t-1}$ is angle opposite to side $({\bf 0},{\bf r}_t)$ so that
\begin{equation}
\cos{\psi_{t,t-1}}=\frac{{\bf v}\cdot{\bf r}_{t-1}}{|{\bf v}|\,|{\bf r}_{t-1}|},
\qquad {\bf v}=\left.\frac{\rmd {\bf r}(\theta)}{\rmd\theta}\right|_{\theta=0},
\end{equation}
using (\ref{geo1}) for ${\bf r}(\theta)$ with ${\bf r}_1={\bf r}_{t-1}$ and
${\bf r}_2={\bf r}_t$.

Evaluating constituents of an admissible ${\cal L}(i_1\ldots i_{N-1},i_N)$,
we see that $d_{t,t-1}=
{\rm dist}_{\mathbb B}(\gamma_{i_1}...\gamma_{i_{t-1}}\gamma_{i_t}[{\bf 0}],
\gamma_{i_1}...\gamma_{i_{t-1}}[{\bf 0}])=
{\rm dist}_{\mathbb B}(\gamma_{i_t}[{\bf 0}],{\bf 0})$, and we have
\begin{equation}
\ell={\rm dist}_{\mathbb B}({\bf 0},\gamma_i[{\bf 0}])\simeq1.762747;
\qquad i=\overline{1,6}.
\end{equation}

Besides, for sufficiently large ${\rm dist}_{\mathbb B}({\bf 0},{\bf r}_t)$ and
${\rm dist}_{\mathbb B}({\bf 0},{\bf r}_{t-1})$ we reduce (\ref{trr}) to
\begin{equation}\label{dd1}
{\rm dist}_{\mathbb B}({\bf 0},{\bf r}_t)\simeq{\rm dist}_{\mathbb B}({\bf 0},{\bf r}_{t-1})
+\ln{(\cosh{\ell}+\sinh{\ell}\,\cos{\psi_{t,t-1}})},
\end{equation}
where the logarithmic term depends on unknown angle $\psi_{t,t-1}$.

In principle, the angle $\psi_{t,t-1}$ depends on ${\bf r}_{t-1}$ (and the action of
$\gamma_{i_t}$), that is, the walk history generated by $\gamma_{i_1},\ldots\gamma_{i_{t-1}}$.
Appealing to the Markov approximation~\cite{Knill} implies neglecting this dependence of
the angle on the point. If we replace the logarithmic term in (\ref{dd1}) with some
$\xi_{i_t,i_{t-1}}$, we get the additive chain:
\begin{eqnarray}
&&{\rm dist}_{\mathbb B}({\bf 0},\gamma_{i_1}[{\bf 0}])=\ell,\quad
{\rm dist}_{\mathbb B}({\bf 0},\gamma_{i_1}\gamma_{i_2}[{\bf 0}])\simeq\ell+\xi_{i_2,i_1},
\nonumber\\
&&{\rm dist}_{\mathbb B}({\bf 0},\gamma_{i_1}\gamma_{i_2}\gamma_{i_3}[{\bf 0}])\simeq\ell
+\xi_{i_3,i_2}+\xi_{i_2,i_1},\ \ {\rm etc.}
\end{eqnarray}

Combining, we arrive at the approximate expression for $p_N(i_1,\ldots i_N)=1$:
\begin{equation}
{\cal L}^*(i_1\ldots i_{N-1},i_N)=N\ell+\sum\limits_{t=1}^{N-1} (N-t)\,\xi_{i_{t+1},i_{t}},
\end{equation}
which depends on the constant $6\times6$ matrix $\|\xi_{i,j}\|$.

Thus, the ${\cal L}$'s spectrum characteristics can be evaluated as
\begin{eqnarray}
&&\left\{ \begin{array}{c}
{\cal L}^*_{\min}(N)\\ {\cal L}^*_{\max}(N)\\ \overline{\cal L}^{\,*}_N
\end{array}\right\}=N\ell+\frac{N(N-1)}{2} \left\{ \begin{array}{c}
\min\limits_{p_2(i,j)=1}(\xi_{i,j})\\
\max\limits_{p_2(i,j)=1}(\xi_{i,j})\\
\overline{\xi}
\end{array}\right\},\label{MarkL}\\
&&(\sigma^*_N)^2=a_0(N) \left[(\overline{\xi^2})_0-\overline{\xi}^{\,2}\right]
+2\sum\limits_{t=1}^{N-2} a_t(N) \left[(\overline{\xi^2})_t-\overline{\xi}^{\,2}\right];
\label{MarkS}\\
&&\overline{\xi}=\frac{1}{v(2)}\sum\limits_{i,j=1}^6 p_2(i,j)\, \xi_{i,j},
\label{aver}\\
&&(\overline{\xi^2})_t=\frac{1}{v(2+t)} \sum\limits_{i_1=1}^6\ldots\sum\limits_{i_{2+t}=1}^6
p_{2+t}(i_1,\ldots,i_{2+t})\,\xi_{i_2,i_1}\,\xi_{i_{2+t},i_{1+t}},
\label{aver2}\\
&&a_t(N)=\sum\limits_{s=1}^{N-1-t} s (s+t),\ \
a_0(N)+2\sum\limits_{t=1}^{N-2} a_t(N)=\frac{N^2(N-1)^2}{4}.
\end{eqnarray}
We see that (\ref{aver2}) takes into account long-range correlations. In fact,
$(\overline{\xi^2})_t\to\overline{\xi}^{\,2}$, and $a_t(N)$ 
decreases for $t\geq3$. It allows us to reduce (\ref{MarkS}) according
to the random walk ideology.

It seems natural to introduce $\|\xi_{i,j}\|$ as
\begin{equation}\label{xi1}
\xi_{i,j}=\sum\limits_{g\in G} \rho_g(i,j)\,
[{\rm dist}_{\mathbb{B}}({\bf 0},g\gamma_j\gamma_i[{\bf 0}])
-{\rm dist}_{\mathbb{B}}({\bf 0},g\gamma_j[{\bf 0}])],
\end{equation}
where $G\subset\Gamma_{\rm BRC}$ is a some (finite) set of generators $g$;
$\rho_g$ is the weight for each $g$.

Note that (\ref{xi1}) generalizes the expression from \cite{Naz},
where $g={\rm id}$ is used in a two-dimensional model. Here,
the reason for involving additional $g$ is to obtain sufficiently large
values of the hyperbolic distance when
$\cosh,\,\sinh\to\sfrac{1}{2}\exp$ in~(\ref{trr}).

\begin{figure}[htbp]
\begin{center}
\includegraphics[width=8cm,angle=0]{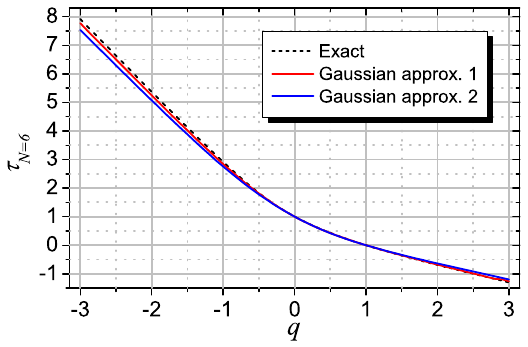}
\end{center}
\vspace*{-4mm}
\caption{\label{fig6}\small Scaling exponent $\tau_N(q)$ for $N=6$ and its approximations.
Both approximations (colored lines) are based on the Gaussian form (\ref{Zapp})
for different sets of parameters: the red line uses the exact characteristics
${\cal L}_{\min}$, ${\cal L}_{\max}$, $\overline{\cal L}_N$, and $\sigma_N$,
while the blue line uses the parameters (\ref{MarkL}), (\ref{MarkS}) in
the Markov approximation.}
\end{figure}

Here we test the implementation, when $G$ is just the generating set (\ref{gami}):
\begin{equation}\label{xi2}
\xi_{i,j}=\frac{1}{5}\,\sum\limits_{k=1}^6 p_2(j,k)\,
[{\rm dist}_{\mathbb{B}}({\bf 0},\gamma_k\gamma_j\gamma_i[{\bf 0}])
-{\rm dist}_{\mathbb{B}}({\bf 0},\gamma_k\gamma_j[{\bf 0}])].
\end{equation}
Applying this definition for $N=6$ gives us ${\cal L}^*_{\min}(N)\simeq12.8784$,
${\cal L}^*_{\max}(N)\simeq33.1276$, $\overline{\cal L}^{\,*}_N\simeq24.6803$,
and $\sigma^{*}_N\simeq2.85126$, while the exact parameters are
${\cal L}_{\min}(N)\simeq13.0367$,
${\cal L}_{\max}(N)\simeq33.0990$, $\overline{\cal L}_N\simeq25.0977$,
and $\sigma_N\simeq3.13207$. By substituting these two sets into (\ref{Zapp})
and calculating $\tau_N(q)$ using (\ref{tD1}), we are able to compare
the approximations for relatively small $|q|$ in Fig.~\ref{fig6}.
Physically, the situation corresponds to the high-temperature mean-field
approximation, since $q\sim 1/T$. Although the double approximation worsens the
description, it eliminates the need to calculate the ${\cal L}$'s spectrum.
We leave possible improvements for further research. Some aspects regarding
the ergodicity of the model may also be clarified.

\section{\label{Sec4}Deformations and Teichm\"uller space}

Let us determine the deformation space ${\rm Def}(M)$ for  
orbifold~$M\simeq\mathbb{B}/\Gamma_{\rm BRC}$ by using the Mostow's generalized
rigidity theorem~\cite{Th78}. Then the deformation space of hyperbolic three-manifold~$M$
may be defined as
\begin{equation}
{\rm Def}(M)\simeq{\rm Teich}(\partial M),
\end{equation}
where ${\rm Teich}(\partial M)$ is the Teichm\"uller space of the marked
structures on the closed and oriented surface $\partial M$.

A crucial step towards our goal is the description of the Teichm\"uller
space ${\rm Teich}(\Sigma)$ associated with the octahedral surface~$\Sigma$
having six cusps and $S_4\times\mathbb{Z}_2$ symmetry,
as shown in Fig.~\ref{fig2}(right). Marking $\Sigma$ and 
stabilizing each of the ideal vertices of $\Sigma$ with the Abelian
subgroup $A_i=\langle \gamma_i,\,{\tilde\gamma}_i\in\Gamma_{\rm BRC}|
 \gamma^n_i{\tilde\gamma}^m_i={\tilde\gamma}^m_i\gamma^n_i\rangle$ formed
by the meridian and longitude for $i=\overline{1,6}$, we first build the
decorated Teichm\"uller space ${\cal T}(\Sigma)$~\cite{Penn87}, that is,
the space of hyperbolic cusp metrics on $\Sigma$ with the addition of
horospheres centered at the cusps. Removing decoration implies projection 
${\cal T}(\Sigma)\mapsto{\rm Teich}(\Sigma)$.

Using the radius-vectors ${\bf e}_{\pm x}=(\pm1\ 0\ 0)^\top$,
${\bf e}_{\pm y}=(0\ \pm1\ 0)^\top$, and ${\bf e}_{\pm t}=(0\ 0\ \pm1)^\top$
for the cusps, let us define the vertex set 
$V=\{{\bf e}_{\pm x},{\bf e}_{\pm y},{\bf e}_{\pm t}\}$ with the cardinality
$|V|=6$, and the geodesic edge set $E=\{E_{ji}\}$ with $|E|=12$, appropriately
connecting the vertices ${\bf e}_j,\,{\bf e}_i\in V$ of the octahedron.
Hereafter, we denote the triangular faces as $F_i$ (or $\Delta_i$) with
evident $|F|=8$.

Decorating the octahedron by means of the set of horospheres
$H=\{H_{\pm x}$, $H_{\pm y}$, $H_{\pm t}\}$ centered at the cusps $V$, all
discrete metrics (finite lengths of edge segments bounded by $H$) form a
manifold of real dimension $|E|$. This manifold is fibered by the discrete
conformal classes representing submanifolds of dimension $|V|$. Besides,
conformal equivalence of the metric sets also admits conformality of
triangulations of the octahedron faces. Thus, the discrete conformal class
correspond to a point in the Teichm\"uller space~${\cal T}_{|V|}$ endowed
with the mapping class group $\pi_0({\rm Aut}(\Sigma))$, which is isomorphic
to the braid group of vertex permutations.

\subsection{Decorating the octahedral surface}

\begin{figure}[htbp]
\begin{center}
\includegraphics[width=4cm,angle=0]{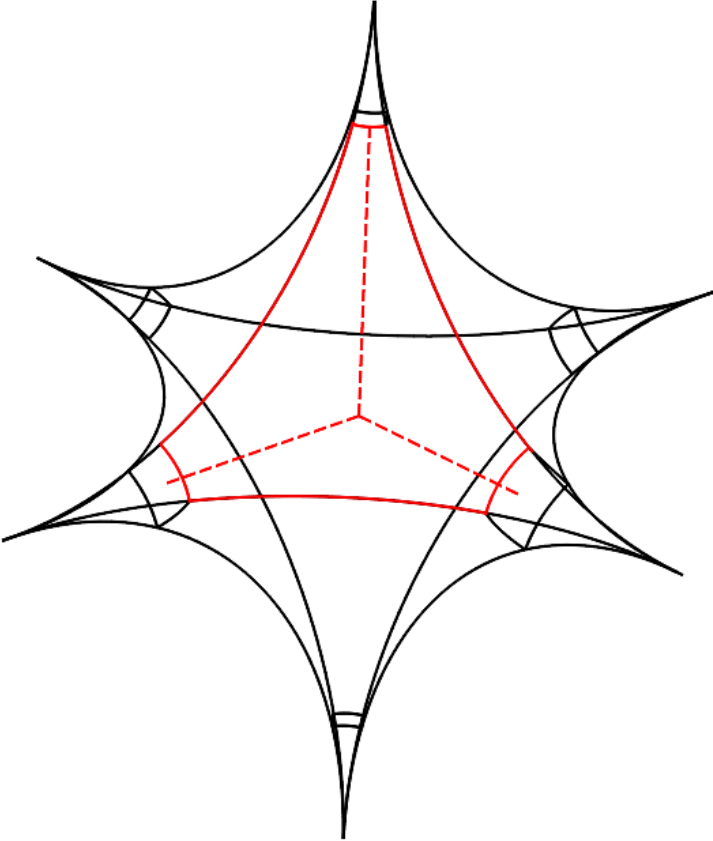}
\end{center}
\vspace*{-4mm}
\caption{\label{fig7}\small Decorated octahedral surface. Intersecting the octahedral
surface, each horosphere $H_i$, centered at octahedron ideal vertex ${\bf e}_i$,
produces a square with right angles. Arbitrarily choosing the heights $h_i$, that is,
the Euclidean distances from the horospheres to the origin (red dashed lines), we
obtain on each octahedron face a hexagon similar to the one formed from the red curves.
The lengths of the red segments of the octahedron edges are equal to $\lambda_{i,j}$,
while the length of each edge of the square contour around the cuspidal tail is
determined by $w_i$.}
\end{figure}

We start to describe the decorated geometry in terms of the set of Euclidean
heights\footnote{We hope that the same notation $h_i$ for generators in Sec.~2
and for heights here will not cause confusion due to the context.}
$h=\{h_{\pm x}, h_{\pm y}, h_{\pm t}\}$, $h_i\in[0;1]$, determining the location
of the horospheres $H$ so that $h_i {\bf e}_i$ is the point of $H_i$ closest to
the origin (see Fig.~\ref{fig7}). We obtain that the endpoint of the 
$E_{ji}$-edge segment terminating at $H_i$ is specified by the radius-vector:
\begin{equation}
{\bf e}_{j,i}=2\,\frac{(1-h_i)^2}{(1-h_i)^2+4}\,{\bf e}_j
+\frac{(1+h_i)^2}{(1-h_i)^2+4}\,{\bf e}_i,
\end{equation}
which tends to ${\bf e}_i\in\partial\mathbb{B}$ when $h_i\to1$.
The other end of the same edge segment is given by ${\bf e}_{i,j}$.
Then, for two adjacent vertices $i$, $j$ and the corresponding heights
$h_i$, $h_j$ we have that the signed hyperbolic length between
${\bf e}_{j,i}$ and ${\bf e}_{i,j}$ is
\begin{eqnarray}
\rho(h_i,h_j)&=&2\,{\rm arcsinh}\,\frac{3(h_i+h_j)-h_ih_j-1}{2\sqrt{2(1-h_i^2)(1-h_j^2)}}.
\end{eqnarray}
Note that if the horospheres overlap, $\rho(h_i,h_j)$ becomes negative.

Thus, $\lambda_{i,j}=\rho(h_i,h_j)$ produce the set $\Lambda=\{\lambda_{\pm x,\pm y},
\lambda_{\pm x,\pm t}, \lambda_{\pm y,\pm t}\}\in\mathbb{R}^{|E|}$ of $\lambda$-lengths
(need not be distinct) of the deco\-rated octahedron edges. These $\Lambda$s serve as
coordinates of Teichm\"uller space. However, the number of independent parameters
remains equal to $|V|$, and they can be further considered as moduli.

Note that the vector tangent to the edge $E_{ji}$ at the endpoint ${\bf e}_{j,i}$ is
\begin{equation}
{\bf v}_{j,i}=(3-h_i)(1+h_i)\,{\bf e}_i-4(1-h_i)\,{\bf e}_j.
\end{equation}

We also need the lengths $w_{i}$ of non-empty paths $F_{j}\cap H_i$  connecting
the nearby ends ${\bf e}_{j_{\alpha},i}$ of the reduced edges $E_{j_{\alpha}i}$.
It is expected that 
$w_i\geq{\rm dist}_{\mathbb{B}}({\bf e}_{j_{1},i},{\bf e}_{j_{2},i})$,
where the ends ${\bf e}_{j_{1},i},{\bf e}_{j_{2},i}\in F_{j}\cap H_i$,
and $w_i\to0$ at $h_i\to1$.

Thus, the length $w_i=w(h_i)$ can be obtained by directly integrating
$\rmd s^{\ }_{\mathbb{B}}$ along the parametrized path between
the nearby ends belonging to $H_i$ and is given by
\begin{equation}
w(h)=\sqrt{2}\,\frac{1-h}{1+h},
\end{equation}
see Appendix~\ref{App2} for details.

Therefore, decoration actually results in applying a right-angled hexagon to
each face $F_i$ of the octahedron (see Fig.~\ref{fig7}), using two
sets of lengths: $\{\lambda_{i_2,i_1},\lambda_{i_3,i_1},\lambda_{i_2,i_3}\}$
and $\{w_{i_1},w_{i_2},w_{i_3}\}$. Since the right angles of hexagons are preserved
for an arbitrary height set~$h$, the decoration has conformal property. As argued
in \cite{BPS15,GL18,Pap17}, we may additionally complicate the structure by
triangulating the hexagons.

Indeed, function $w(h)$ is the subject of the identity:
\begin{equation}\label{wcon}
w(h)=w({\tilde h})\,\rme^{\epsilon\tau_{\bf e}},\quad
\tau_{\bf e}={\rm dist}_{\mathbb{B}}(h{\bf e},{\tilde h}{\bf e}),\quad
\epsilon={\rm sign}({\tilde h}-h),\quad
{\bf e}\in V.
\end{equation}
This relation induces the boost of the height $h$ for vertex ${\bf e}$:
\begin{equation}
{\tilde h}=\frac{h+\tanh{(u_{\bf e}/2)}}{1+h \tanh{(u_{\bf e}/2)}},
\end{equation}
where $u_{\bf e}\in\mathbb{R}$ and should be equal to $\epsilon\tau_{\bf e}$
to reproduce (\ref{wcon}).

Additional relations appear when introducing the midpoint
of the edge $E_{ji}$, i.e.
${\bf m}_{j,i}=(1-2^{-1/2})({\bf e}_j+{\bf e}_i)$, 
and the distances determined by $h_i$ and $h_j$:
\begin{equation}
p_{j,i}={\rm dist}_{\mathbb{B}}({\bf m}_{j,i},{\bf e}_{j,i}),\qquad
p_{i,j}={\rm dist}_{\mathbb{B}}({\bf m}_{j,i},{\bf e}_{i,j}),
\end{equation}
so that $\lambda_{j,i}=p_{j,i}+p_{i,j}$. We have that $w_i=\exp{(-p_{j,i})}$ and
$w_j=\exp{(-p_{i,j})}$ \cite{Penn87}.

For ideal triangle $\Delta_i$ (octahedron face $F_i$) with vertices
$\{{\bf e}_{i_1},{\bf e}_{i_2},{\bf e}_{i_3}\}$ decorated by horospheres
$\{H_{i_1},H_{i_2},H_{i_3}\}$, we are focusing on the configurations
corresponding to the triangle inequalities:
\begin{equation}
\lambda_{i_2,i_1}+\lambda_{i_3,i_1}>\lambda_{i_2,i_3},\quad
\lambda_{i_2,i_1}+\lambda_{i_2,i_3}>\lambda_{i_3,i_1},\quad
\lambda_{i_3,i_1}+\lambda_{i_2,i_3}>\lambda_{i_2,i_1}.
\end{equation}
Denoting $p_{i_2,i_1}=p_{i_3,i_1}\equiv a_i$, $p_{i_1,i_2}=p_{i_3,i_2}\equiv b_i$,
and $p_{i_2,i_3}=p_{i_1,i_3}\equiv c_i$, one has $\lambda_{i_2,i_1}=a_i+b_i$,
$\lambda_{i_3,i_1}=a_i+c_i$, and $\lambda_{i_2,i_3}=b_i+c_i$. We see that \cite{BPS15}
\begin{eqnarray}
w_{i_1}=\rme^{-a_i},&\quad a_i=\frac{1}{2}(\lambda_{i_2,i_1}+\lambda_{i_3,i_1}-\lambda_{i_2,i_3}),&
\nonumber\\
w_{i_2}=\rme^{-b_i},&\quad b_i=\frac{1}{2}(\lambda_{i_2,i_1}+\lambda_{i_2,i_3}-\lambda_{i_3,i_1}),&
\\
w_{i_3}=\rme^{-c_i},&\quad c_i=\frac{1}{2}(\lambda_{i_3,i_1}+\lambda_{i_2,i_3}-\lambda_{i_2,i_1}).&
\nonumber
\end{eqnarray}
Conversely, we claim that $\lambda_{i,j}=-\ln{(w_i w_j)}$.

Parameterizing $h_i=\tanh{(u_i/2)}$ by ${\bar u}_i=u_i-\frac{1}{2}\ln{2}$
for $i=\overline{1,|V|}$, the discrete conformal metrics are
determined as
\begin{equation}
w_i=\rme^{-{\bar u}_i},\qquad
\ell_{i,j}=\rme^{\lambda_{i,j}/2},\qquad
\lambda_{i,j}={\bar u}_i+{\bar u}_j,
\end{equation}
where $\ell_{i,j}$ and $w_i$ are the Penner's lengths and heights \cite{Penn87}.
Then, two extended sets $\{\lambda^{(1)}_{i,j},w^{(1)}_i\}$ and 
$\{\lambda^{(2)}_{i,j},w^{(2)}_i\}$ belong to the same discrete conformal class
and are {\it isometric} if there exists a one-parameter M\"obius transformation:
\begin{equation}
h^{(2)}_i=\frac{h^{(1)}_i+\tanh{(\tau/2)}}{1+h^{(1)}_i \tanh{(\tau/2)}},\quad
{\bar u}^{(2)}_i={\bar u}^{(1)}_i+\tau,\quad
 i=\overline{1,|V|}.
\end{equation}

Thus, an ordered tuple $U_0=\{{\bar u}_{\pm x},{\bar u}_{\pm y},{\bar u}_{\pm t}\}$
of six fixed positive numbers produces the equivalence class $\{U_{\tau}\}$
preserving a marking, where each set
$U_{\tau}=\{{\bar u}_{\pm x}+\tau,{\bar u}_{\pm y}+\tau,{\bar u}_{\pm t}+\tau\}$ has
non-negative components for $\tau\in\mathbb{R}$.

Conformal stretching of the surface structure can be generated by
a convex ``energy'' function ${\cal E}(\{ {\bar u}_i\})$ so that
\begin{equation}
\frac{\rmd {\bar u}_i}{\rmd\tau}=-\frac{\partial {\cal E}(\{ {\bar u}_i\})}{\partial{\bar u}_i}.
\end{equation}
This equation is similar to the discrete Ricci/Yamabe flow \cite{BPS15,GL18},
and the volume of cuspidal tails truncated by horospheres can serve as 
${\cal E}(\{ {\bar u}_i\})$. As an alternative,
other functions are also proposed in \cite{BPS15,GL18}.

\subsection{The mapping class group}

Some permutations of ${\bar u}_i$ in $U_0$, which lead to an inequivalent
$\widetilde{U}$, could preserve the face hexagons constructed on the base
of $U_0$. They are generated by two fourth-order rotations about the 
orthogonal axis, which determine the {\it mapping class group}
${\cal G}^*=\langle\hat\tau_{1},\hat\tau_2\,|\,\hat\tau_{1}^4=\hat\tau_2^4=1\rangle$
\cite{McM13}. Representatives of ${\cal G}^*$ rearrange the vertices of
the cyclic boundaries defined as follows.

Let $\{\Delta_{\alpha}\,|\,\alpha=\overline{1,4}\}$ be a cycle of adjacent triangles
(octahedron faces) so that $\Delta_{\alpha}\cap\Delta_{\alpha+1}=E_{\alpha}$, where
we regard the index $\alpha$ as cyclic, $\Delta_{5}=\Delta_{1}$. Denoting the ideal
edges of $\Delta_{\alpha}$ as $\{E_{\alpha-1},E_{\alpha},\widetilde{E}_{\alpha}\}$,
the collection $\{\widetilde{E}_{\alpha}\,|\,\alpha=\overline{1,4}\}$ forms
boundary of the cycle $\{\Delta_{\alpha}\,|\,\alpha=\overline{1,4}\}$. Obviously,
in the planes $\mathbb{D}_{t=0}$, $\mathbb{D}_{y=0}$, $\mathbb{D}_{x=0}$ there are
three four-sided boundaries separating three pairs of triangle cycles.

Operating by $\hat\tau_k$, four heights ${\bar u}_i$ are cyclically permuted
in one of such planes, while the rest two heights lying on the axis orthogonal
to the plane are fixed. Thereby we determine the {\it Dehn twists} in three
orthogonal planes:
\begin{eqnarray}
&&\hat\tau_1[U_0]=\{{\bar u}_{\mp y},{\bar u}_{\pm x},{\bar u}_{\pm t}\},\qquad
\hat\tau_1^{-1}[U_0]=\{{\bar u}_{\pm y},{\bar u}_{\mp x},{\bar u}_{\pm t}\},
\nonumber\\
&&\hat\tau_2[U_0]=\{{\bar u}_{\mp t},{\bar u}_{\pm y},{\bar u}_{\pm x}\},\qquad
\hat\tau_2^{-1}[U_0]=\{{\bar u}_{\pm t},{\bar u}_{\pm y},{\bar u}_{\mp x}\},\\
&&\hat\tau_3[U_0]=\{{\bar u}_{\pm x},{\bar u}_{\mp t},{\bar u}_{\pm y}\},\qquad
\hat\tau_3^{-1}[U_0]=\{{\bar u}_{\pm x},{\bar u}_{\pm t},{\bar u}_{\mp y}\},
\nonumber
\end{eqnarray}
where $\hat\tau_3=\hat\tau_1^{-1}\hat\tau_2^{-1}\hat\tau_1$, and
$\hat\tau_1\hat\tau_2\hat\tau_1=\hat\tau_2\hat\tau_1\hat\tau_2$ from the braid group $B_3$.

It means that any two different markings of the octahedral surface are related
by the action of the mapping class group ${\cal G}^*$. We immediately deduce
that the {\it moduli space} is ${\cal M}^*(\Sigma)\simeq\mathbb{R}^{|V|}_+/{\cal G}^*$.

Note that the mapping class group, acting combinatorially on six (unmarked)
punctures of the sphere $\partial\mathbb{B}$, is associated to the braid group~$B_6$
\cite{Farb}. This could be also considered as the starting point of our reasoning
\cite{McM13}.

To formalize the action of ${\cal G}^*$, we introduce the six-dimensional vector
\begin{equation}
{\bf u}=\left(\footnotesize{\begin{array}{c} 1\\ 0\end{array}}\right)\otimes{\bf u}_+
+\left(\footnotesize{\begin{array}{c} 0\\ 1\end{array}}\right)\otimes{\bf u}_-,\quad
{\bf u}_{\pm}=\left(\footnotesize{\begin{array}{c} {\bar u}_{\pm x}\\
{\bar u}_{\pm y}\\ {\bar u}_{\pm t}\end{array}}\right).
\end{equation}

Thus, the implementation $\rho: {\cal G}^*\mapsto {\rm Mat}(6,\{0,1\})$ is
given by the matrices of permutations $T_i=\rho(\hat\tau_i)$, expanded due
to the Kronecker multiplication as
\begin{eqnarray}
&T_i=I_2\otimes M_i+J_2\otimes N_i,&\\
&I_2=\left(\footnotesize{\begin{array}{cc} 1& 0\\ 0& 1\end{array}}\right),\qquad
J_2=\left(\footnotesize{\begin{array}{cc} 0& 1\\ 1& 0\end{array}}\right),&
\end{eqnarray}
\begin{eqnarray}
M_1=\left(\footnotesize{\begin{array}{ccc} 0& 0& 0\\ 1& 0& 0\\ 0& 0& 1\end{array}}\right),\quad
&M_2=\left(\footnotesize{\begin{array}{ccc} 0& 0& 0\\ 0& 1& 0\\ 1& 0& 0\end{array}}\right),&\quad
M_3=\left(\footnotesize{\begin{array}{ccc} 1& 0& 0\\ 0& 0& 0\\ 0& 1& 0\end{array}}\right),\\
N_1=\left(\footnotesize{\begin{array}{ccc} 0& 1& 0\\ 0& 0& 0\\ 0& 0& 0\end{array}}\right),\quad
&N_2=\left(\footnotesize{\begin{array}{ccc} 0& 0& 1\\ 0& 0& 0\\ 0& 0& 0\end{array}}\right),&\quad
N_3=\left(\footnotesize{\begin{array}{ccc} 0& 0& 0\\ 0& 0& 1\\ 0& 0& 0\end{array}}\right),
\end{eqnarray}
so that $T_i^4=I_6$ and $\det{T_i}=-1$; $I_n$ is the unit matrix in $n$ dimensions.

One has $\rho(g_1g_2)=\rho(g_1)\rho(g_2)$ for $g_{1,2}\in{\cal G}^*$, and
\begin{eqnarray}
T^{-1}_i&=&I_2\otimes M^\top_i+J_2\otimes N^\top_i,\\
T_iT_j&=&I_2\otimes (M_iM_j+N_iN_j)+J_2\otimes (M_iN_j+N_iM_j),\\
T_i{\bf u}&=&\left(\footnotesize{\begin{array}{c} 1\\ 0\end{array}}\right)\otimes
(M_i{\bf u}_++N_i{\bf u}_-)
+\left(\footnotesize{\begin{array}{c} 0\\ 1\end{array}}\right)\otimes
(M_i{\bf u}_-+N_i{\bf u}_+),
\end{eqnarray}
where $J_2^2=I_2$ and $N_iN_j=0$ by construction.

The new configuration is then determined by $\widetilde{{\bf u}}=\rho(g)[{\bf u}]$,
$g\in{\cal G}^*$. Obviously, the permutations preserve the form:
\begin{equation}
\xi_n=\sum\limits_{s={x,y,t}}({\bar u}^n_{+s}+{\bar u}^n_{-s}),\quad
n\in\mathbb{N}.
\end{equation}

Setting $J=T_1T_2$ and $S=T_1 J$ results in the relation $S^2=J^3=I_6$ inherent
in $PSL(2,\mathbb{Z})\simeq\mathbb{Z}_2*\mathbb{Z}_3$, and $B_3$ looks like
the central extension of $PSL(2,\mathbb{Z})$~\cite{ARab}. We also note
the $q$-deformed analogue of $B_3$, presented and discussed in \cite{AKos}.

To describe the group structure, we represent $T_i$, $i=\overline{1,3}$,
in the form:
\begin{equation}
T_i=I_2\otimes R_i+(I_2+J_2)\otimes N_i,\qquad
R_i=M_i-N_i.
\end{equation}

We find that each $R_i=\exp{\left(\frac{\pi}{2}\,X_i\right)}$
is generated by $X_i$ from  $so(3)$ algebra:
\begin{eqnarray}
&X_1=
\left(
\footnotesize{
\begin{array}{rrr}
0& -1& 0\\
1& 0& 0\\
0& 0& 0
\end{array}}
\right),\quad
X_2=
\left(
\footnotesize{
\begin{array}{rrr}
0& 0& -1\\
0& 0& 0\\
1& 0& 0
\end{array}}
\right),\quad
X_3=
\left(
\footnotesize{
\begin{array}{rrr}
0& 0& 0\\
0& 0& -1\\
0& 1& 0
\end{array}}
\right),&\\
&[X_i,X_j]=\varepsilon_{ijk} X_k,\quad \{i,j,k\}=\{1,2,3\},&
\end{eqnarray}
where $\varepsilon_{ijk}$ is the Levi-Civita tensor, and the Casimir
operator is then equal to
\begin{equation}
X_1^2+X_2^2+X_3^2=-2 I_3.
\end{equation}

Further, introducing
${\cal R}_i=I_2\otimes R_i=\exp{\left(\frac{\pi}{2}\,I_2\otimes X_i\right)}$,
we decompose $T_i$ as
\begin{equation}
T_i={\cal R}_i {\cal A}_i,\qquad
{\cal A}_i=I_2\otimes(I_3+R^{-1}_iN_i)+J_2\otimes(R^{-1}_iN_i),
\end{equation}
where ${\cal A}_i^2=I_6$ indicates the second-order cyclic relation.

Excluding $\langle {\cal A}_1, {\cal A}_2\rangle$
from ${\cal G}^*$, we are left with the group
$\langle R_1, R_2\,|\,R_1^4=R_2^4=I_3\rangle$ in $SO(3)$, which
permutes $\alpha_s$ in initial vector
$\boldsymbol{\alpha}=(\alpha_x\ \alpha_y\ \alpha_t)^\top\in\mathbb{R}^3$,
when identifying
\begin{equation}
\alpha_s\leftrightarrow {\bar u}_{\pm s},\quad 
-\alpha_s\leftrightarrow {\bar u}_{\mp s},\quad
s=\{x,y,t\}.
\end{equation}

This completes the analysis of the structure of the mapping class group
${\cal G}^*$ operating in Euclidean (moduli) space~\cite{Farb} and leaving
the following interval invariant:
\begin{equation}\label{Rmm}
\rmd s^2_{{\cal M}}=\sum\limits_{s=x,y,t}(\rmd{\bar u}_{+s}^2+\rmd{\bar u}_{-s}^2).
\end{equation}

On the other hand, assigning positive real numbers to the octahedron ideal edges 
($E_{ij}\mapsto\lambda_{i,j}$) and thereby defining the point $P\in{\cal T}(\Sigma)$
in Teichm\"uller space, then $\lambda$-lengths are natural for the action of
the mapping class group ${\cal G}(\Sigma)$. Thus, ${\cal G}(\Sigma)$
acts on ${\cal T}(\Sigma)$. If $\varphi\in{\cal G}(\Sigma)$, there is
a map $\varphi_*:{\cal T}(\Sigma)\mapsto{\cal T}(\Sigma)$,
which for an arc $E$ gives $\lambda(E;P)=\lambda(\varphi E;\varphi_* P)$.
Since the point $T$ is determined by $\Lambda$-set, it is easy to show that
$\varphi_*=\varphi^{-1}$. Having already described the action of the mapping class
group ${\cal G}^*$, we omit here explicit transformations of $\lambda$-lengths,
which become identities using $U$ and $g[U]$, where $g\in{\cal G}^*$.

Thus, when formulating physical models, it is possible to take into account the conformality
of surface structures and their transformations under the action of mapping class
group. One of the attractive aspects of further study is the geometry quantization.

\subsection{Towards area quantization}

Consider a star-shaped body bounded by the octahedron edges and formed by
three ideal squares lying in orthogonal planes $\mathbb{D}_{t=0}$,
$\mathbb{D}_{y=0}$, $\mathbb{D}_{x=0}$. Let us analyze how to approach a quantum spectrum
of the area of such squares.

\begin{figure}[htbp]
\begin{center}
\includegraphics[width=4cm,angle=0]{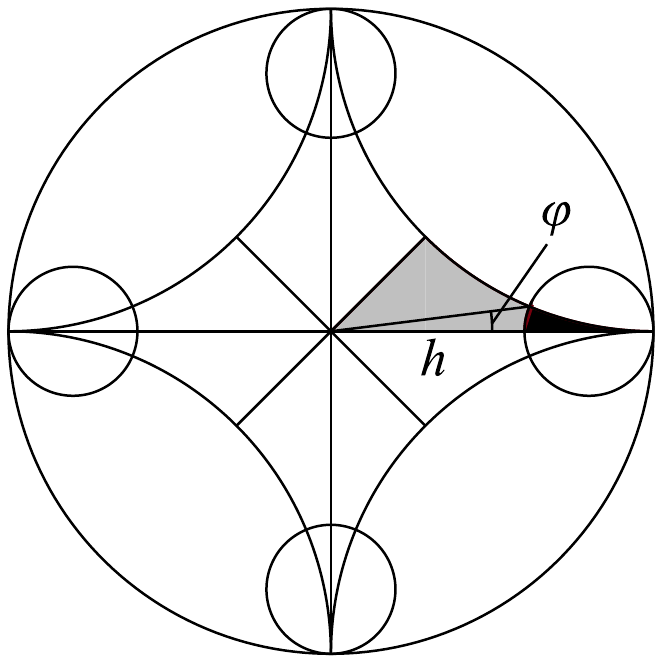}
\end{center}
\vspace*{-4mm}
\caption{\label{fig8}\small Decorated ideal square divided into eight parts.
The gray segment area $A_{\Delta}$ is given by Eq.~(\ref{Ad}), where the
defect $\pi/4-A_{\Delta}$ determines the area of black cuspidal tail.}
\end{figure}

Dissecting the one decorated square into eight parts (see Fig.~\ref{fig8}),
we obtain that the area of each part depends on the height $h$ as
\begin{equation}\label{Ad}
A_{\Delta}(h)=\frac{\pi}{4}-\frac{1-h}{1+h}.
\end{equation}
For simplicity, we set $h_i=h\in[2^{-1/2};1]$ for $i=\{\pm x, \pm y, \pm t\}$
and write down the total area of the right-angled octagon as $A=8A_{\Delta}$.

Regarding $h=\tanh{(u/2)}$ as above, we obtain the geometric relations:
\begin{equation}\label{ww}
\frac{1-h}{1+h}=\rme^{-u}=\sqrt{\frac{1}{2}\,\tan{\varphi}},
\end{equation}
where the angle $\varphi$ is indicated in Fig.~\ref{fig8}.
Note that (\ref{ww}) is exactly equal to the hyperbolic length of
horocircle segment being the fourth edge of the gray region.

This immediately leads to the (stationary) antikink:
\begin{equation}\label{phi1}
\varphi({\bar u})=\arctan{\rme^{-2 {\bar u}}},\qquad
{\bar u}\in[0;+\infty),
\end{equation}
where the module ${\bar u}=u-\frac{1}{2}\ln{2}$ is such that introduced earlier.

Thus, (\ref{phi1}) is a subject of a one-dimensional sine-Gordon model with the
following action integral and equation of motion \cite{Man04}:
\begin{eqnarray}
&&S=\int\left\{\frac{1}{2}\,(\partial_{\bar u}\varphi)^2
+\frac{1}{4}\,\left[1-\cos{(4\varphi)}\right]\right\} \rmd {\bar u},\\
&&\partial^2_{\bar u}\varphi-\sin{(4\varphi)}=0,\label{eom1}
\end{eqnarray}
which are invariant under the global transformation $\varphi\to\varphi+\pi n/2$,
$n\in\mathbb{N}$. In the general case of distinct heights $h_i$, $S$
is being evidently written for a sextet of fields $\{\varphi_i\,|\,i=\overline{1,|V|},
|V|=6\}$, preserving invariance under the permutations induced by the mapping class group.
In the absence of interaction between the sextet components, the equations of
motion take the form ({\ref{eom1}}), but differ in the initial conditions.

The energy of (\ref{phi1}) is determined by the integral:
\begin{equation}
\int_0^{\infty}\left\{(\partial_{\bar u}\varphi)^2
+\frac{1}{2}\,\left[1-\cos{(4\varphi)}\right]\right\} \rmd {\bar u}=1.
\end{equation}
%

We assume that our problem concerns the quantization of the relation:
\begin{equation}
2\left(\frac{\pi}{4}-A_{\Delta}\right)^2=\tan{\varphi},
\end{equation}
where the right-hand side reduces to the form in term of the self-action $W$:
\begin{equation}
\tan{\varphi}=\frac{\sqrt{W(\varphi)}}{1+\sqrt{1-W(\varphi)}},
\qquad
W(\varphi)=\frac{1}{2}\,\left[1-\cos{(4\varphi)}\right].
\end{equation}
We leave the search for a solution and exploration to future work that needs to
show possible connection with low-dimensional loop quantum gravity and spin
foam~\cite{Rovelli14,RS95}. A similar task would arise when quantizing lengths
(see also \cite{Naz05}).

\begin{figure}[htbp]
\begin{center}
\includegraphics[width=5cm,angle=0]{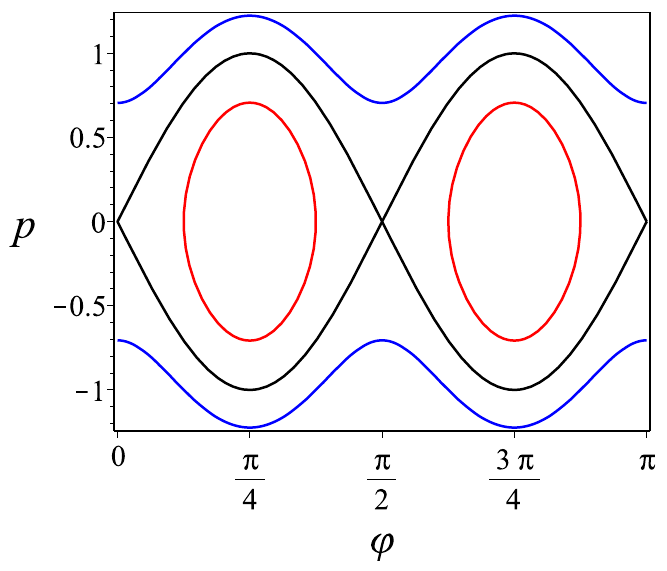}
\end{center}
\vspace*{-4mm}
\caption{\label{fig9}\small Phase portrait of trajectories at fixed energy $H$
in (\ref{Ham1}). The black curve corresponds to $H=0$. Blue curve is typical
for $H>0$ (the scattering), while the red curve is for $H<0$. At $H=-1/2$,
the ellipses are contracted into points with the coordinates $((2n+1)\pi/4;0)$,
$n\in\mathbb{Z}$.}
\end{figure}

Finally, we would like to analyze the area of phase space.
Introducing the canonical momentum $p=\partial_{\bar u}\varphi$, inducing
the Poisson bracket $\{\varphi,p\}=1$, and taking into account
that $W(\varphi)=\sin^2{(2\varphi)}$, the dynamics is generated by the Hamiltonian:
\begin{equation}\label{Ham1}
H=\frac{p^2-\sin^2{(2\varphi)}}{2}.
\end{equation}
The regimes of such a model are depicted in Fig.~\ref{fig9}.

The solution (\ref{phi1}) corresponds to $H=0$ and serves as a boundary
for the mode $H<0$ with cyclic trajectories. Denoting $\varepsilon=-2H\geq0$
in these cases, let us define the action variable associated with the
``quantum'' of phase-space area:
\begin{eqnarray}
J&\equiv&\oint p(\varphi)\,\rmd\varphi
=-4\,\int_{\varphi_0(\varepsilon)}^{\pi/4} \sqrt{W(\varphi)-\varepsilon}\,\rmd\varphi\\
&=&2 E(\sqrt{1-\varepsilon})-2 \varepsilon K(\sqrt{1-\varepsilon}),
\end{eqnarray}
where $\varphi_0(\varepsilon)=(1/2) \arcsin{\sqrt{\varepsilon}}$;
$K(k)$ and $E(k)$ are the complete elliptic integrals of the first and
second kind, respectively \cite{AbSt}. Hence, the area of $n\in\mathbb{N}$
(red or black) cycles at $\varepsilon\in[0;1)$ is equal to $n J(\varepsilon)$,
where $J(0)=2$ and $J(1)=0$.

Of course, the obtained results are easily generalized for the case of
distinct heights $h_i$, when the induced symplectic 2-form becomes 
$\omega=\sum_{i=\overline{1,6}} \rmd p_i\wedge\rmd\varphi_i$.
This is seen as a tool for constructing an algebra whose generator
spectra solve the quantization problem~\cite{Naz13,Hurt}.

\section{Discussion}

Embedding all the structures associated with the Borromean rings
complement (BRC) into the Poincar\'e unit ball $\mathbb{B}$ with its isometry
group ${\rm Isom}(\mathbb{B})$, we begin with a description of the BRC geometry
and finding the realization $\Gamma_{\rm BRC}\subset{\rm Isom}(\mathbb{B})$ of
the fundamental group $\pi_1({\rm BRC})\simeq\mathbb{Z}_2^3\rtimes {\cal C}_3$.
Using Thurston's proposal regarding the BRC tessellation of two octahedra~\cite{Th78},
we do this in several steps.

We found it convenient to first fix the four-vertex basis of one octahedron in
the plane of unit disc $\mathbb{D}$ by using the parabolic $SU(1,1)$-isometries
(see Fig.~\ref{fig2}). Having extended the action of the parabolic generators
up to three dimensions due to transformation from~\cite{Mat06} and further
operating with rotations around the main axes, we obtain the overfull set of
twelve generators. Namely, we arrive at six pairs of parabolic generators
$\langle h_i, \tilde{h}_i\rangle$ of longitudes and meridians, which fix
each of the six vertices of the octahedron and form Abelian subgroups
(see Appendix~\ref{App1}). 

Next, we choose three pairs of generators (\ref{subG}) related to distinct
axes. They correspond to the torus $T^3$ and are isomorphic to the BRC
fundamental group~\cite{Abe13}. To obtain $\Gamma_{\rm BRC}$ in the Wirtinger
representation~(\ref{gRel}), we express three of the six generators in terms
of the remaining ones (\ref{gBRC}), chosen as independent.

Thus, we first get twelve generators operating in three dimensions, and then
reduce their number to six and finally to three. This allows us to analyze
the structure of $\Gamma_{\rm BRC}$ and to apply it in geometry and constructing
Teichm\"uller space. Indeed, twelve non-independent generators are convenient
for tiling $\mathbb{B}$ with octahedra, as well as for obtaining the maximally
symmetric complement~\cite{Ad20} represented by the rhombic dodecahedron in
Fig.~\ref{fig3}(right). Perhaps the applied scheme will also be useful for
implementing the group in terms of quaternions~\cite{Wil93}.

Having realized the group $\Gamma_{\rm BRC}$, it is natural to turn to the
problem of obtaining functions that are invariant under the action of the group.
In fact, it was constructed in~\cite{Mat06} a functional basis consisting of
Jacobi $\theta$-functions symmetrized with respect to the BRC group as a
subgroup of $SL(2,\mathbb{C})$.

Here, we consider the Cayley tree rooted at the origin and embedded in the ball
$\mathbb{B}$ for the group $\Gamma_{\rm BRC}$ and focus on the multifractal
properties of the partition function ${\cal Z}_N(q)$ from (\ref{Z1}) defined for
a truncated tree up to $N$th generation. It is viewed as a kind of Poincar\'e
series of the Boltzmann weights for discrete paths, where the moment order~$q$
corresponds to the inverse temperature $1/T$ in physics and takes positive and
negative values. Since the conventional Poincar\'e series of Boltzmann weights,
determined by the only hyperbolic distance of order $N$, between the vertices
of the tree, does not exhibit multifractality, we use the functional ${\cal L}$
with an ``interaction'' of the form (\ref{L1}) and order~$N^2$. This is
similar to the perimetric characteristic of the surface formed by adjacent
triangles connecting the root point and two nearby vertices \cite{Naz}.
Besides, the six-branch graphs that emerged there can imitate dendritic
polymers~\cite{dp21}.

Numerical analysis revealed fast stochastization of ${\cal L}$ with increasing $N$
and multifractal behavior of ${\cal Z}_N(q)$ (see Fig.~\ref{fig4}), when the exponents
of its characteristics are simply compared for different models and admit a physical
interpretation~\cite{Fed}. 

Thus, the obtained multifractal exponents, such as the fractal dimensions,
indicate the absence of a dominant subset of paths and the need to take into
account all graphs, at least approximately, in the desired analytical 
description. On the other hand, for large~$N$ the behavior becomes conditioned by
the central limit theorem. This allowed us to apply Markov chains (random walk)
and calculate the partition function in the Gaussian approximation (mean field
approximation in physics), valid for relatively small~$|q|$ (see Fig.~\ref{fig6}).
Note that for a better analytical description it was necessary to improve the
formula for transition matrix (\ref{xi2}), the introduction of which eliminates
the need to numerically calculate the ${\cal L}$-spectrum for all graphs.

Touching upon the concept of the deformation space in the last part of our work,
we appeal to the generalized Mostow's theorem~\cite{Th78}. Thus, the key point 
to describing the deformations of the quotient $\mathbb{B}/\Gamma_{\rm BRC}$
is the decorated Teichm\"uller space of conformal structures on the regular
octahedral surface with ideal vertices (cusps). Here we follow Penner's
receipt~\cite{Penn87}, although there is its generalization in \cite{Luo14}.

Decorating the octahedral surface involves incorporating horospheres with the centers
at the vertices and obtaining curves of their intersection with octahedron faces,
which always remain orthogonal to the octahedron edges. Conformality is justified
by maintaining right angles regardless of the size of each horosphere, which is
regulated by boost~\cite{Penn87,BPS15,GL18}.

According to Penner~\cite{Penn87}, the decorated Teichm\"uller space, when a surface
{\it is marked} by a particular realization of the fundamental group, consists of sets
of finite hyperbolic lengths of edge segments bounded by horospheres. It is due to
the fact that the octahedron faces are ideal triangles, i.e. the octahedral surface
is initially triangulated. And we may introduce the manifold $\mathbb{R}_+^{|E|}$ for
the number of edges $|E|=12$.

However, by fibering, information is encoded in
${\cal T}_{|V|}\simeq\mathbb{R}_+^{|V|}$ at $|V|=6$ that is the number of vertices.
Geometrically, this means that the set of six positive real numbers (heights), which
are the distances from the origin to the horopsheres in the direction of the vertices,
defines all the elements of the global structure (see Fig.~\ref{fig7}).
Two such sets are isometrically equivalent and belong to the same {\it equivalence
class} if they are related by a one-parameter M\"obius transform (boost). Besides,
the ordered set of six hyperbolic lengths is the object of the action of the
mapping class group ${\cal G}^*$.

To maintain the conformal structure induced by the decoration on the octahedral
surface, the group ${\cal G}^*$ acts on the ordered set by cyclic permutations,
which are generated by two fourth-order rotations~\cite{McM13}.
We are implementing ${\cal G}^*$ in Sec.~4.2, describing its structure, and
reducing it to a subgroup of $SO(3)$. On the other hand, ${\cal G}^*$
can be considered as a permutation group of punctures on $\partial\mathbb{B}$
and associated with the
braid group~\cite{Farb}. Anyway, we arrive at the {\it moduli space}
$\mathbb{R}_+^{|V|}/{\cal G}^*$ with the Euclidean metric, omitting the operation
of the mapping class group in $\mathbb{R}_+^{|E|}$.

With the goal to quantize geometry of the BRC, we were faced with the need to induce
a differential two-form and introduce the algebra of observables \cite{Naz05,Naz13}.
In principle, it can be done by means of dynamical model and its symplectic
form~\cite{Hurt}. Fortunately, by analyzing the area of the decorated square in
Fig.~\ref{fig8}, we geometrically related the angular size $\varphi$ of the curve that
simultaneously belongs to the horosphere and the face, and the hyperbolic distance
${\bar u}$ from the origin to it. Thereby we arrived at the kink (\ref{phi1}) of
the one-dimensional sine-Gordon model~\cite{Man04}. Then, passing to the Hamiltonian
formalism, the posed problem is resolved in the cotangent bundle. Additionally, using
a phase portrait of the model and the action variable, we also took a step towards
quantizing the area of the phase space. Although a detailed study of the desired
algebra still remains a prospect for the future.

\section*{Acknowledgments}

A.V.N. thanks to A.M.~Gavrilik (BITP) for inspiring discussions and acknow\-ledges
support from the National Academy of Sciences of Ukraine (by its project No.~0122U000888)
and the Simons Foundation.

\appendix

\section{\label{App1}Three-dimensional transformations}

Using the extension (\ref{gext}) preserving $\rmd s^2_{\mathbb{B}}$,
the generators of $\Gamma$ induce the group $\Gamma_*$ of coordinate transformations
(parabolic isometries) in three-dimensional ball $\mathbb{B}$:
\begin{eqnarray}
&&\hspace{-9mm}
{\footnotesize
h^{n}_1\left[\left(\begin{array}{c} x\\ y\\ t \end{array}\right)\right]=
\begin{array}{c} 1 \\ \overline{n^2[t^2+(y-1)^2]+(nx-1)^2} \end{array}
\left(\begin{array}{c}
 x-n[x^2+(y-1)^2+t^2]\\
 y+n^2[x^2+(y-1)^2+t^2]-2nx\\
 t
\end{array}\right)
},\label{A1}\\
&&\hspace{-9mm}
{\footnotesize
h^{n}_2\left[\left(\begin{array}{c} x\\ y\\ t \end{array}\right)\right]=
\begin{array}{c} 1 \\ \overline{n^2[t^2+(x+1)^2]+(ny-1)^2} \end{array}
\left(\begin{array}{c}
 x-n^2[(x+1)^2+y^2+t^2]+2ny\\
 y-n[(x+1)^2+y^2+t^2]\\
 t
\end{array}\right)
},\\
&&\hspace{-9mm}
{\footnotesize
h^{n}_3\left[\left(\begin{array}{c} x\\ y\\ t \end{array}\right)\right]=
\begin{array}{c} 1 \\ \overline{n^2[t^2+(y+1)^2]+(nx+1)^2} \end{array}
\left(\begin{array}{c}
 x+n[x^2+(y+1)^2+t^2]\\
 y-n^2[x^2+(y+1)^2+t^2]-2nx\\
 t
\end{array}\right)
},\\
&&\hspace{-9mm}
{\footnotesize
h^{n}_4\left[\left(\begin{array}{c} x\\ y\\ t \end{array}\right)\right]=
\begin{array}{c} 1 \\ \overline{n^2[t^2+(x-1)^2]+(ny+1)^2} \end{array}
\left(\begin{array}{c}
 x+n^2[(x-1)^2+y^2+t^2]+2ny\\
 y+n[(x-1)^2+y^2+t^2]\\
 t
\end{array}\right)
}.
\end{eqnarray}
In the plane of $\mathbb{D}$ at $t=0$,
they reduce to linear-fractional transformations $h^n_k[z]$.

To tile $\mathbb{B}$ by octahedra, it needs to extend the group $\Gamma_*$ by
new (parabolic) generators which act in the planes orthogonal to the disc
$\mathbb{D}$:
\begin{eqnarray}
&&\hspace{-9mm}
{\footnotesize
h^{n}_{+}\left[\left(\begin{array}{c} x\\ y\\ t \end{array}\right)\right]=
\begin{array}{c} 1 \\ \overline{n^2[y^2+(t-1)^2]+(nx+1)^2} \end{array}
\left(\begin{array}{c}
x+n[x^2+y^2+(t-1)^2]\\
y\\
t+n^2[x^2+y^2+(t-1)^2]+2nx
\end{array}
\right)
},\\
&&\hspace{-9mm}
{\footnotesize
{\tilde h}^{n}_{+}\left[\left(\begin{array}{c} x\\ y\\ t \end{array}\right)\right]=
\begin{array}{c} 1 \\ \overline{n^2[x^2+(t-1)^2]+(ny+1)^2} \end{array}
\left(\begin{array}{c}
 x\\
 y+n[x^2+y^2+(t-1)^2]\\
 t+n^2[x^2+y^2+(t-1)^2]+2ny
\end{array}\right)
},\\
&&\hspace{-9mm}
{\footnotesize
h^{n}_{-}\left[\left(\begin{array}{c} x\\ y\\ t \end{array}\right)\right]=
\begin{array}{c} 1 \\ \overline{n^2[y^2+(t+1)^2]+(nx+1)^2} \end{array}
\left(\begin{array}{c}
 x+n[x^2+y^2+(t+1)^2]\\
 y\\
 t-n^2[x^2+y^2+(t+1)^2]-2nx
\end{array}\right)
},\\
&&\hspace{-9mm}
{\footnotesize
{\tilde h}^{n}_{-}\left[\left(\begin{array}{c} x\\ y\\ t \end{array}\right)\right]=
\begin{array}{c} 1 \\ \overline{n^2[x^2+(t+1)^2]+(ny+1)^2} \end{array}
\left(\begin{array}{c}
 x\\
 y+n[x^2+y^2+(t+1)^2]\\
 t-n^2[x^2+y^2+(t+1)^2]-2ny
\end{array}\right)
}.\label{A8}
\end{eqnarray}

Note that $h_{+}$ and ${\tilde h}_{+}$ (as well as $h_{-}$ and
${\tilde h}_{-}$) refer to the same fixed point. It means that
their composition is also a parabolic generator, and
${\tilde h}^n_{\pm}\,h^m_{\pm}=h^m_{\pm}\,{\tilde h}^n_{\pm}$. Looking for 
${\tilde h}_k$, $k=\overline{1,4}$, to create analogous Abelian subgroups,
it needs to rotate $h_{1,3}$ by the angle $\pm\pi/2$ around basis vector ${\bf j}$
and rotate $h_{2,4}$ around vector ${\bf i}$ by using (\ref{RRot}). This gives
us a pair of generators corresponding to the meridian and longitude at each vertex.

\section{\label{App2}Calculating $w(h)$}

\begin{figure}[htbp]
\begin{center}
\includegraphics[width=5cm,angle=0]{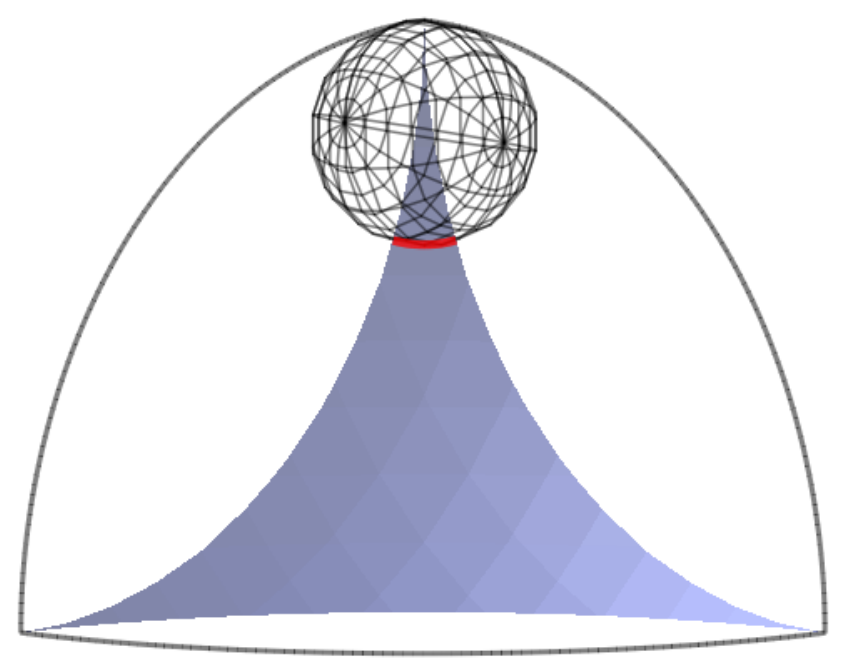}
\end{center}
\vspace*{-4mm}
\caption{\label{fig10}\small The intersection (in red) of octahedral face and horosphere.}
\end{figure}

Let us calculate the hyperbolic length $w(h)$ of the red curve $C$ in Fig.~\ref{fig10}
resulted from intersection of the octahedral face $(x-1)^2+(y-1)^2+(t-1)^2=2$ for
$0\leq x,y,t<1$ and the horosphere $x^2+y^2+[t-(1+h)/2]^2=(1-h)^2/4$ for the cusp
$(0\ 0\ 1)^\top$. The endpoints of $C$ are
\begin{equation}
\left(0,\ \frac{2(1-h)^2}{(1-h)^2+4},\ \frac{(1+h)^2}{(1-h)^2+4} \right)^\top,\quad
\left(\frac{2(1-h)^2}{(1-h)^2+4},\ 0,\ \frac{(1+h)^2}{(1-h)^2+4} \right)^\top.
\end{equation}

Parametrizing $C$ by $x\in D$ as
\begin{eqnarray}
&&t(x)=\frac{3-2x(1-h)+h^2-2\sqrt{(1-h)^2(1+2x-x^2)-8x^2}}{(1-h)^2+4},\\
&&y(x)=\frac{1-h}{2}[1-t(x)]-x,\qquad
D=\left[0; \frac{2(1-h)^2}{(1-h)^2+4}\right],\nonumber
\end{eqnarray}
one has that
\begin{equation}
w(h)\equiv2\int_D\frac{\sqrt{1+[y^\prime(x)]^2+[t^\prime(x)]^2}}{1-x^2-y^2(x)-t^2(x)}\,\rmd x
=\sqrt{2}\,\frac{1-h}{1+h}.
\end{equation}



\begin{thebibliography}{0}

\bibitem{Th78}
W.P.~Thurston, The geometry and topology of 3-manifolds. {\it Princeton University Lecture
Notes} (1978).

\bibitem{W78}
N.~Wielenberg, The structure of certain subgroups of the Picard group.
{\it Math. Proc. Camb. Phil. Soc.} {\bf 84}, 427 (1978).

\bibitem{Mat06}
K.~Matsumoto, Automorphic functions with respect to the
fundamental group of the complement of the Borromean rings.
{\it J. Math. Sci. Univ. Tokyo} {\bf 13}, 1–11 (2006).

\bibitem{Abe13}
R.~Abe and I.R.~Aitchison, Geometry and Markoff’s spectrum for $\mathbb{Q}(\rmi)$, I.
{\it Transact. AMS} {\bf 365}(11), 6065–6102 (2013).

\bibitem{Penn87}
R.C.~Penner, The decorated Teichm\"uller space of punctured surfaces.
{\it Comm. Math. Phys.} {\bf 113}, 299-339 (1987).

\bibitem{Kauff02}
L.H.~Kauffman and S.J.~Lomonaco, Quantum entanglement and topological entanglement.
{\it New J. Phys.} {\bf 4}, 73 (2002).

\bibitem{Iq24}
M.~Iqbal, N.~Tantivasadakarn, R.~Verresen {\it et al.} Non-Abelian topological
order and anyons on a trapped-ion processor. {\it Nature} {\bf 626}, 505–511
(2024).

\bibitem{Grimm06}
T.~Kraemer, M.~Mark, P.~Waldburger {\it et al.} Evidence for Efimov quantum
states in an ultracold gas of caesium atoms. {\it Nature} {\bf 440}, 315–318 (2006).

\bibitem{BrH06}
E.~Braaten and H.-W.~Hammer, Universality in few-body systems with
large scattering length. {\it Phys. Rep.} {\bf 428}, 259-390 (2006).

\bibitem{Ch04}
K.S.~Chichak {\it et al.} Molecular Borromean Rings. {\it Science} {\bf 304},
1308-1312 (2004).

\bibitem{Rovelli14}
C.~Rovelli and F.~Vidotto, In: {\it Covariant Loop Quantum Gravity: An Elementary
Introduction to Quantum Gravity and Spinfoam Theory} (Cambridge University Press,
Cambridge, 2014).

\bibitem{Kas}
C.~Kassel, {\it Quantum groups} (Springer-Verlag, New York, 1995).

\bibitem{dp21}
D.A.~Tomalia, J.B.~Christensen and U.~Boas, {\it Dendrimers, Dendrons and Dendritic
Polymers. Discovery, Applications and the Future} (Cambridge University Press, Cambridge, 2012).

\bibitem{Fed}
J.~Feder, {\it Fractals} (Plenum Press, New York, 1988).

\bibitem{BPS15}
A.I.~Bobenko, U.~Pinkall, and B.A.~ Springborn, Discrete conformal maps and ideal
hyperbolic polyhedra. {\it Geom. Topol.} {\bf 19}, 2155-2215 (2015).

\bibitem{GL18}
X.D.~Gu, F.~Luo, J.~Sun, and T.~Wu, A discrete uniformization theorem for
polyhedral surfaces. {\it J. Differential Geom.} {\bf 109}, 223-256 (2018).

\bibitem{Man04}
N.~Manton and P.~Sutcliffe, {\it Topological Solitons}
(Cambridge University Press, Cambridge, 2004).

\bibitem{Naz05}
A.~Nazarenko, Time level splitting in quantum Chern–Simons gravity.
{\it Class. Quantum Grav.} {\bf 22}, 2107-2120 (2005).

\bibitem{Naz13}
A.V.~Nazarenko, Area quantization of the parameter space of Riemann surface
in genus two. {\it Ukr. J. Phys.} {\bf 58}, 1055-1064 (2013).

\bibitem{Hurt}
N.E.~Hurt, {\it Geometric Quantization in Action: Applications of Harmonic
Analysis in Quantum Statistical Mechanics and Quantum Field Theory}
(D. Reidel Publishing Company, 1983).

\bibitem{MNY05}
K.~Matsumoto, H.~Nishi, and M.~Yoshida, Automorphic functions for the
Whitehead-link-complement group. {\it Kyushu University Preprint Series
in Mathematics} (2005).

\bibitem{Rod}
O.~Rodrigues, {\it Journal de Math\'ematiques Pures et Appliqu\'ees de Liouville}
{\bf 5}, 380–440 (1840).

\bibitem{HNN}
G.~Higman, B.H.~Neumann, and H.~Neumann, Embedding theorems for groups.
{\it J. London Math. Soc.} {\bf 24}, 247–254 (1949).

\bibitem{Wil93}
J.B.~Wilker, The quaternion formalism for M\"obius groups in four or
fewer dimensions. {\it Lin. Alg. Appl. } {\bf 190}, 99-136 (1993).

\bibitem{Ad20}
C.~Adams, A.~Calderon, and N.~Mayer, Generalized bipyramids and hyperbolic
volumes of alternating $k$-uniform tiling links.
{\it Topol. Appl.} {\bf 271}, 107045 (2020).

\bibitem{RT21}
J.G.~Ratcliffe and S.T.~Tschantz, Cusp transitivity in hyperbolic 3-manifolds.
{\it Geom. Dedicata} {\bf 212}, 141–152 (2021).

\bibitem{Hoff22}
N.R.~Hoffman, Cusp types of quotients of hyperbolic knot complements.
{\it Proc. Amer. Math. Soc. Ser. B} {\bf 9}, 336-350 (2022).

\bibitem{Naz}
A.V.~Nazarenko, Directed random walk on the lattices of genus two.
{\it Int. J. Mod. Phys. B} {\bf 25}, 3415–3433 (2011).

\bibitem{Knill}
O.~Knill, {\it Probability and stochastic processes with applications}
(Overseas Press, New Delhi, 2009).

\bibitem{AbSt}
M.~Abramowitz and I.A.~Stegun (eds), {\it Handbook of mathematical functions with
formulas, graphs, and mathematical tables} (Dover Publications, New York, 1972).

\bibitem{Pap17}
A.~Papadopoulos and S. Yamada, Deforming hyperbolic hexagons with applications
to the arc and the Thurston metrics on Teichm\"uller spaces.
{\it Monatsh. Math.} {\bf 182}, 913–939 (2017).

\bibitem{McM13}
C.T.~McMullen, Braid groups and Hodge theory. {\it Math. Ann.} {\bf 355},
893–946 (2013).

\bibitem{Farb}
B.~Farb and D.~Margalit, {\it A Primer on Mapping Class Groups}
(Princeton University Press, Princeton, 2012).

\bibitem{ARab}
S.~Albeverio and S.~Rabanovich, On a class of unitary representations of
the braid groups $B_3$ and $B_4$. {\it Bul. Sci. Math.} {\bf 153},
35-56 (2019).

\bibitem{AKos}
S.~Albeverio and A.~Kosyak, $q$-Pascal's triangle and irreducible
representations of the braid group $B_3$ in arbitrary dimension.
ArXiv: 0803.2778 [math.QA].

\bibitem{RS95}
C.~Rovelli and L.~Smolin, Discreteness of area and volume in quantum gravity.
{\it Nucl. Phys. B} {\bf 442}, 593-619 (1995).

\bibitem{Luo14}
F.~Luo, Rigidity of polyhedral surfaces, I. {\it J. Differential Geom.} {\bf 96},
241-302 (2014).

\end{thebibliography}
\end{document}